\documentclass{aastex6}

\usepackage{graphicx}
\usepackage[utf8]{inputenc}
\usepackage{hyperref}
\usepackage{multirow}
\usepackage{amsmath}
\usepackage{multirow}

\begin{document}

\title{Classifying Radio Galaxies with Convolutional Neural Network}
\author{A. K. Aniyan}
\affil{Department of Physics and Electronics, Rhodes University, Grahamstown, South Africa}
\affil{SKA South Africa, $3^{rd}$ Floor, The Park, Cape Town, South Africa}
\and
\author{K.Thorat}
\affil{Department of Physics and Electronics, Rhodes University, Grahamstown, South Africa}
\affil{SKA South Africa, $3^{rd}$ Floor, The Park, Cape Town, South Africa}

\begin{abstract}
We present the application of deep machine learning technique to classify radio images of extended sources on a morphological basis using convolutional neural networks. In this study, we have taken the case of Fanaroff-Riley (FR) class of radio galaxies as well as radio galaxies with bent-tailed morphology. We have used archival data from the  Very Large Array (VLA) - Faint Images of the Radio Sky at Twenty Centimeters (FIRST) survey and existing visually classified samples available in literature to train a neural network for morphological classification of these categories of  radio sources. Our training sample size for each of these categories is $\sim$200 sources, which has been augmented by rotated versions of the same. Our study shows that convolutional neural networks can classify images of the FRI and FRII and bent-tailed radio galaxies with high accuracy (maximum precision at 95\%) using well-defined samples and  “fusion classifier”, which combines the results of binary classifications, while allowing for a mechanism to find sources with unusual morphologies. The individual precision is highest for bent-tailed radio galaxies at 95\% and is 91\% and 75\% for the FRI and FRII classes, respectively, whereas the recall is highest for FRI and FRIIs at 91\% each, while bent-tailed class has a recall of 79\%. These results show that our results are comparable to that of manual classification while being much faster. Finally, we discuss the computational and data-related challenges associated with morphological classification of radio galaxies with convolutional neural networks.\\

\end{abstract}


\section{Introduction}
\label{inroduction}
With the advent of the Square Kilometer Array (SKA) radio telescope  along with its precursor facilities, we expect the radio sky to be surveyed at high speed and to unprecedented sensitivity. While this may enable paradigm shifts in the studies of radio sources, it comes with very high data volumes. For example, the typical image size from the MeerKAT telescope is estimated to be 11.13 TB \footnote{MeerKat SDP group, S. Ratcliffe, B. Merry, Bennett, T.} (individual MeerKat surveys expect to deal with a large number of objects, e.g. MeerKlass survey expects to find upwards of 200000 radio sources in the HI emission line \footnote{Józsa, G.I.G., SKA South Africa, private communication}). This  introduces challenges in all the steps of data reduction from RFI mitigation to calibration and imaging. The science level data volume is expected to be similarly formidable. For example, extrapolating from the Square Kilometre Array Design Studies (SKADS) \citep{skads_2010}, gives a source density of $6.2 \times 10^4$
sources per square degree for a survey reaching 1 microJy at 1.4 GHz \citep{padovani_rev}, and the Evolutionary Map of the Universe survey (EMU) with Australian Square Kilometer Array Pathfinder (ASKAP) \citep{norris_emu_2011} is expected to find about 70 million radio sources, while covering two-thirds of the sky. Handling this amount of data is not possible through manual studies: automation of data processing is therefore essential. 

In this study, we consider the case of radio galaxy classification in the image domain. Traditionally, source detection and classification is trivially done for unresolved and slightly resolved radio sources through various source finding software (these radio sources may in fact be ``components" rather than true sources, for example, these might be lobes of a double radio galaxy). Somewhat more recently, there have been several attempts to classify/identify radio sources through automated techniques (crowd-sourcing is an alternative, e.g. Radio Galaxy Zoo \citep{RGZ_2015}) such as pattern recognition and decision trees \citep{proctor11},  source matching and pattern recognition \citep{van_velzen_2015}  and self-organizing maps \citep{self_organizing_maps_polsterer}. The latter is an example of a Machine Learning technique, which has come into increasing use in the recent years (especially in pulsar and transient detection/identification, see \citep{morello2014,Eatough_2010,bates_2012,wagstaff_2016}).  

The typical process of source detection and classification hinges on using source finding software to generate source component catalogs. These components need to be combined (or identified) as a single source, where necessary (source de-blending would be the opposite issue). This is especially important for extended sources, which are more likely to get divided in multiple components.  These can be AGN-powered or star-forming galaxies. In the past, studies have tended to classify these sources by visual examination. This quickly grows impracticable with increasing survey sizes. Here, we consider the application of deep machine learning techniques to classify extended extragalactic radio sources, more specifically, AGN-powered radio galaxies.

Machine learning  methods have been applied to a variety of astronomical problems, such as star-galaxy classification \citep{odewahn1992}, red-shift estimation \citep{benitez2000bayesian}, classification of optical transients \citep{mahabal2011discovery} and unsupervised source segmentation  \citep{hocking2015teaching} among others. These methods have been robust and reliable, and have performed with high accuracy. For example, with classification of stars and galaxies the best available model has over 99\% accuracy \citep{kim2017star}. Estimation of redshifts with machine learning has been done with accuracies over 92\% \citep{cavuoti2017metaphor,cavuoti2017cooperative}. Machine learning methods have been incorporated to real-time transient classification systems and they perform with accuracies of more than 90\% \citep{mahabal2008towards,mahabal2011discovery}. 

Classic machine learning algorithms such as support vector machines (SVM), K-nearest neighbors (KNN) and decision trees generally learn on `features' extracted from the observational data \citep{kotsiantis2007supervised}. Features represent unique characteristics of the raw data and are domain specific. Feature extraction is carefully done so that the chosen features will  represent specific physical properties of the system \citep{guyon2006introduction}. The efficiency of the learning algorithm mainly depends on the quality of the features used\citep{blum1997selection}. Such machine learning algorithms are generally called shallow learning methods  \citep{chen1995machine}. In principle, shallow learning methods \textit{learn} from the features rather than the raw data which may be images or time series values. A good understanding of the data and the objective under investigation is required to properly extract the features and fine tune the machine learning algorithm. The features extracted may also not encapsulate the distinct properties of the object. 

Deep learning is a branch of machine learning in which the machine learning algorithms learn directly from the data instead of features \citep{bengio2009learning}. Deep learning is advantageous in situations where engineered features do not completely capture the physics of the raw data and the machine learning algorithm is not able to learn with minimal loss \citep{arel2010deep,lecun2015deep}. 
 
 Recent developments in computing technology, mainly with graphic processing units (GPU), has accelerated  the development of deep neural networks (DNN) for different applications. The seminal work by  \citet{hinton2006fast} and \citet{bengio2007scaling} made it possible to train DNNs for complex classification and regression problems with very high accuracy. It is interesting to note that DNNs have been beating all other shallow learning algorithms by huge margins \citep{lecun2015deep} especially in applications such as object recognition\citep{krizhevsky2012imagenet}, image captioning \citep{vinyals2015show}, speech recognition \citep{graves2013speech}, video natural language processing\citep{collobert2008unified} and many more. These considerations make DNNs a very useful tool for the classification of extragalactic radio sources performed in this study.

There are a variety of ways in which radio galaxies can be classified. The classification can be made on a purely morphological basis, or can take other parameters into account, e.g. spectral index, host galaxy brightness at optical/infrared wavelengths, host galaxy spectra/type. Restricting ourselves to classification schemes based solely on morphology, we find schemes such as the Fanaroff and Riley classification (FR henceforth) \citep{FR74}, Wide-angled tailed and Narrow-angled tailed radio galaxies etc. In this study, we have chosen to restrict our investigations to classifications made only on the basis of the radio morphology. The advantage of this choice being that the source samples used are not restricted by the availability of ancillary data. In turn, the classification algorithm is valid for data which has no or limited ancillary data available. This is a major consideration for deep radio surveys as well as surveys which are outside the coverage of available ancillary data.

The first classification scheme which we consider here is the Fanaroff-Riley (FR) classification. The FR scheme divides extended radio galaxies in two classes, designated as FRI and FRII, membership of which depends on the ratio R of the distance between the brightest points in the source and the total size of the source.Radio galaxies for which R $ < 0.5$ are classified as FRI and those for which R $ \geq 0.5$ are classified as FRII. Typical features associated with FRI-type radio galaxies include diffuse, plume-like jet(s) and cores which are brighter than jets/lobes. FRII-type radio galaxies on the other hand, show bright `hotspots', typically at the end of the lobes and cores which are less bright than these. The FR classification scheme, which starts on a morphological basis, also corresponds to a division in radio power ($P_{1.4GHz} = 10^{25}$ W/Hz) and possibly host galaxy optical luminosity \citep{OL96}. For a detailed discussion, see \citet{saripalli12}. 

The FRI/II sources form the bulk of AGN-powered radio galaxies. These are important sources of the feedback processes in the cosmic structure formation \citep{croton2006}. Several arguments have been advanced to explain the morphological differences including intrinsic differences in the AGNs powering these sources, the environments of the sources large-scale or galactic scales or the mode of the accretion. However, these factors have not been able to successfully explain the FR dichotomy \citep{Gendre2013}. Apart from these sources, there are the so-called FR 0 sources \citep{Sadler2014,Baldi2016} as well as sources with 'hybrid' morphology \citep{Gopal-Krishna2000} which require further examination. An issue in the latter studies is the relatively small fraction of FRI sources found in current all-sky surveys - due to their relatively high detection and completeness threshold, high redshift and/or low luminosity, low-surface brightness source populations are not probed well. Upcoming all-sky surveys with SKA will probe FRI populations to high redshifts \citep{kapinska2015} and would be able to answer these questions \citep{kharb2016}. 

The other category of sources considered in this work are bent-tail sources. Bent-tailed radio galaxies include Wide Angled Tailed (WAT), Head-tail (HT), Narrow Angled Tailed (NAT) radio galaxies.
As their name suggests, these radio galaxies have jets ('tails') which are bent at angle from the host optical galaxy, the nature of the angle between the jets determining if the radio galaxy is a WAT or NAT. In some of these galaxies the jets are swept back to such an extent that they appear as a head (the core) and a tail. These are the HT radio galaxies. The peculiar radio morphology of the bent-tailed sources is generally attributed to their environment, typically a galaxy cluster or a group \citep{burns98}. As such, these sources can be used as tracers of clusters of galaxies \citep{rgz16}, \citep{mao_atlas}, especially at high redshifts where  the information from optical or X-ray bands may be unavailable or sparse. 

The plan of the paper is as follows. In the next section we describe the source sample chosen for training and classification. Section 3  gives a concise background of Convolutional Neural Networks. Section 4 contains a description of the specific neural network model we have chosen, the pre-processing needed for the sample source images and the training process. Section 5 explains the classification model used to determine the final classification of the sources. Section 6 presents the results and discussion. In section 7, we briefly summarize the study and present conclusions.

\section{Sample Selection}
\label{sample_selection}
In this section we describe the sample formation for this study. We have formed separate samples for FRI, FRII and Bent-tailed radio galaxies respectively. The factors to consider while selecting the samples were high sample numbers and images which are well-resolved as well as the free availability of images. With these constraints, we have decided to restrict ourselves to sources from the Faint Images of Radio Sky at Twenty Centimeter (FIRST) \citep{FIRST1995} radio survey. As described below, since there are no source samples of each category which are sufficiently large, we have combined several different samples of sources, further, creating artificial sources by processing the sources from these samples (see Section ~\ref{preprocessing_sample_images} for details). 

We initially selected the FRI-II sample from a subset of the Combined NVSS and FIRST Galaxies sample \citep{config1,config2} (CoNFIG henceforth). This sample of radio sources was compiled specifically to address the need and lack of samples of FRI-II sources in the literature. The CoNFIG sample was compiled from an overlapping region from the NRAO VLA Sky Survey (NVSS) \citep{NVSS1998} and FIRST surveys. The CoNFIG sample is divided  into four sub-samples of varying flux density limits in NVSS, named CoNFIG-1-4, with $S_{1.4GHz}\geq 1.3,0.8, 0.2 $\& $0.05$ Jy respectively (CoNFIG-2-4 are spatially subsets of CoNFIG-1). It should be noted that even the faintest sources in the sample are bright relative to the the bulk of the sources expected to be detected in upcoming surveys. 

In total, the source catalogue from CoNFIG contains 859 sources. This CoNFIG sample was classified by morphological basis into two categories, FRII and FRI radio galaxies (as well as Compact sources and sources of Uncertain morphology, which we do not include in this study). As the NVSS images for most of these sources are unresolved with the NVSS beam FWHM of $45"$, the structural information is obtained with FIRST images which has a beam FWHM of $5"$. The criteria for the classification were presence of 'hotspots' at the edge of radio lobes as well as alignment of the lobes (if the lobes showed hotspots and were aligned, the source was classified as FRII; collimated jets and hotspots close to the core were taken as signs of FRI radio galaxies, note that this includes bent-tailed radio galaxies). In the present study we make use of only the sources classified as FRI/II from these. The FRI/II radio galaxies have an associated flag which can be understood as the degree of confidence in the classification of the source; the flag can either be 'confirmed' or 'possible'. The final classification of the sample provides 71 FRIs (50 confirmed) and 406 FRIIs (with 390 confirmed). As an initial sample we have chosen the 50 confirmed FRIs and 390 confirmed FRIIs. The sparsity of the FRI-type radio galaxies is due to the relatively shallow flux density limits of the CoNFIG survey. For example, at $z=0.15$, the median redshift of FRI radio galaxies in CoNFIG-4 (which is the deepest and spatially the smallest CoNFIG region), the limiting flux density for the other three regions corresponds to a radio power above the nominal radio power divide between FRI/FRII classes. The bulk of the FRIs comes from low redshifts, while the reverse is true for the FRIIs.     

To supplement the smaller number of the FRI radio galaxies and address the imbalance in the training set (see Section \ref{preprocessing_sample_images} for more details), we decided to include the recent FRICAT catalog of FRI radio galaxie \citep{FRICAT}. The FRICAT catalog is a subsample of the \citet{bestandheckman12} sample, by imposing an upper redshift cut of $z = 0.15$, which gave an initial sample of 3357 sources. A further constraint of the size of the radio emission of atleast 30 kpc from the centre of the host galaxy as seen in FIRST images was applied (corresponding to 11".4 for the most distant objects in the FRICAT catalog -thus giving several resolution elements for the smallest source in the sample). Further, sources displaying only FRI morphology (one sided and two sided jets with as well as narrow-angled tailed objects were included). This classification was done visually by all the three authors independently and a source was included in the catalog if at least two of the authors agreed on the classification. This makes the classification more robust; this is also similar to the procedure we have adopted independently (see Section \ref{classification_model}). Including the FRICAT model gives another 219 FRIs (we have excluded the sample of small FRI galaxies included in the FRICAT catalog in the present study). It should be noted that the majority of the FRI source sample for this study is from low redshift universe, while the majority of FRII radio galaxies corresponds to relatively high redshifts. This also means that for a given physical extent, FRIswould have more structural detail.    

For bent radio galaxies, we have made use of the catalog from \citet{proctor11}, where the FIRST radio source database has been classified along morphological categories using a combination of pattern-recognition tools and visual inspection (the latter for sources with more than four components and thus expected  complex morphology). For details of the classification method, see \citet{proctor03} and \citet{proctor06}. In brief, sources in the FIRST catalog are separated in groups, with low-count groups (those with fewer than 3 members) being classified using decision tree pattern recognition techniques in various categories (WAT, NAT, W-shaped sources etc.), and higher-count groups classified using visual inspection. We  make use of only the latter category to form a sample of bent-tailed sources. These sources have been visually examined and classified into a variety of types. From these, we have chosen only the confirmed WAT and NAT radio galaxies, excluding those sources where the WAT and NAT identification is uncertain (marked by '?' next to the classification in the table). This gave us 299 bent-tailed radio galaxies.
\begin{table}[!htbp]
\centering
\begin{tabular}{lllll}
\hline
\multicolumn{1}{l}{} & \multicolumn{1}{l}{Initial Sample Size} & \multicolumn{1}{l}{Image-based Cut} & \multicolumn{1}{l}{Morphology based Cut} & \multicolumn{1}{l}{Final Sample Size} \\ \hline
FRI Sources            & 269                                      & 77                                   & 14                                        & 178                                    \\  \hline
FRII Sources           & 390                                      & 92                                   & 14                                        & 284                                    \\   \hline
Bent-tailed Sources    & 299                                      & 11                                   & 34                                        & 254                                 \\  \hline 
\end{tabular}
\caption{Table summarizing the sample selection process; the image-based cut refers to sources excluded due to presence of artefacts, lack of structural information or very large source size, the morphology based cut refers to sources discarded due to confusion regarding the morphology. }
\label{table_source_selection}
\end{table}
From these initial samples, we have excluded all the source images in which there are strong artefacts. We have also excluded those images which contain multiple sources, as well as sources too large to fit in the cutout and sources small to have sufficient structural information. This reduces the sample size to 47 FRIs from the CoNFIG sample and 145 FRIs from the FRICAT sample (giving 192 FRIs in total), 298 FRIIs and 288 bent-tailed radio galaxies.   
Since all these are samples based on visual inspection, there is a possibility of confusion in the assigned class of a source due to different studies estimating different morphologies for the same source. To resolve this issue, we have excluded all \textit{cross-matched} sources from the FRI and FRII samples which have been marked as confirmed WAT/NAT, W-shaped or Ring (and Ring-lobe) morphologies in either \citet{config2} or \citet{proctor11}. We have removed the bent-tailed radio galaxies from the FRICAT by visual inspection. After removing these sources, we are left with $178$ FRIs, $284$ FRIIs and $254$ bent-tailed radio galaxies. This process is summarized in Table ~\ref{table_source_selection}. In the next section, we describe the convolutional neural network which will be trained in classification using this sample of sources.

\section{Convolutional Neural Networks}
\label{convolutional_neural_network}
Artificial neural networks (ANN), inspired by biological neurons, try to approximate nonlinear functions from a set of inputs by combination of simple functions \citep{cybenko1989approximation}. ANNs generally consist of a network of interconnected neurons which may have many inputs and a single output like a biological neuron. Having a proper learning rule and activation functions, such interconnected neurons in a specific architecture can be used for classification and regression applications \citep{jain1996artificial}.

The output $\mathit{y}$ of a single neuron can be mathematically be represented as

\begin{equation}\label{neuron}
y = \sum_{j=1}^{d} w_{j}x_{j} + w_{0}
\end{equation}
 where $\mathit{x_{j}}$ are the different inputs to the neuron, $\mathit{w_{j}}$ are the weights to the corresponding inputs and $\mathit{w_{0}}$ is the bias term. 
The $\mathit{w_{j}x_{j}}$ term represents a dot product. The output y is then usually passed through an activation function. Similar to the action potential in a biological neuron which decides the rate of neuron firing, the activation function in an artificial neuron restricts the neuron output to normalizable values.

\begin{equation} \label{neuron-new}
\hat{y} = f \left(y \right)
\end{equation}
 $\mathit{f}$ in equation \ref{neuron-new} is the activation function. The activation functions are of different types namely threshold functions, piece-wise linear functions and sigmoid functions \citep{duda2012pattern}. Similarly large number of neurons can be interconnected with multiple layers of neurons having distinct activation function \citep{duda2012pattern}. Therefore Equation \ref{neuron} can be rewritten as

\begin{equation}\label{hidden}
Net_{k} =  \sum_{j} ^{n_H} y_{j} w_{kj} + w_{k0}
\end{equation}

In equation \ref{hidden}, $\mathit{k}$ are each units in the output layer and $\mathit{n_H}$ are the number of hidden layers. Combinations of such $\mathit{n_{H}}$ layers can be used to learn non-linear functions with backpropagation \citep{hecht1989theory}. During the learning process, the inputs, multiplied with their associated weight and bias propagate from the input layer to the output layer through the different hidden layers of neurons. This is commonly referred to as forward pass or forward propagation. At the output the error between the calculated output and expected output is estimated and this error is sent back from the output to the input layer to adjust the weights of the neurons. This is called backward pass or backpropagation. 

In a convolutional neural network, the dot product in equation \ref{hidden} is replaced by a convolution operator. Hence, $\mathit{w_{j}}$ will be a vector instead of a single value as in the case of a normal neural network. $\mathit{w_{j}}$ is often called a kernel or filter. This facilitates convolutional neural networks to directly operate on raw data such as images or time series data as opposed to feature vectors in normal neural networks \citep{lecun1995convolutional}.  

Yann LeCun for the first time showed the successful application of convolutional neural network (CNN)  to digit recognition \citep{lecun1995convolutional}. CNNs have been widely used for image classification \citep{lawrence1997face}, speech signal processing \citep{hinton2012deep} and text classification \citep{collobert2008unified}. 

CNNs are also referred to as Time Delay Neural Networks (TDNN) because they are generally insensitive to translations of a pattern \citep{duda2012pattern}. This property is achieved by a method called weight sharing, which constrains backpropagation to generate the same weight values for similar a region in the input space. The input space refers to the data space which is input to the network, for example images or time series data. Weight sharing is an important property of CNNs which allows the generation and extraction of translation independent features from the raw data. It can be explained with the following example. Consider a cutout image of an FRI type galaxy. The galaxy remains FRI even if the same galaxy is at the center of the image cutout or even at any of the corners provided it is clearly visible. The same is the case when the spatial size of the galaxy within the image changes. There are specific properties that make the galaxy FRI type irrespective of its position, size,flipping or mirroring and tilt. The CNN learns features that are shift, translational and rotational invariant through weight sharing. This simply means that the set of weight values which represent the FRI galaxy features are same irrespective of translational and rotational variations across different samples of the same type. Thus the weights are shared for a specific property of a class are shared among different samples. For example the set of weights which extract features for the two hot spots of FRII galaxy are the same for any sample of FRII type galaxy.
CNNs also have the same feedforward operations as that of a conventional neural network, enabling application of similar learning principles.

\begin{figure}[!ht]
	\centering
	\includegraphics[scale=1]{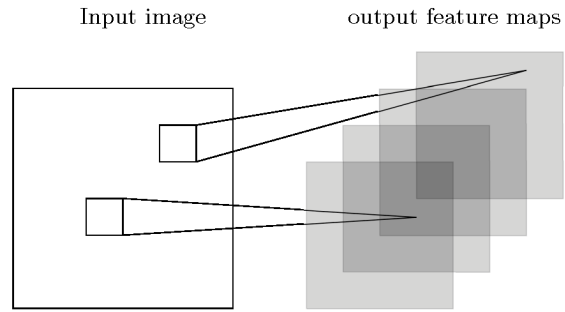}
	\caption{Illustration of a single convolutional layer with multiple output feature maps. With a single input it is possible to learn different features with different filters/kernels, thus creating a depth of feature maps. The gray squares represent different filters learned in a single layer. The small squares on the input image with pointers to the feature maps simply shows the convolution kernel sliding across the image to generate values in the feature maps.}
	 \label{cnn}
\end{figure}

One of the main advantages of convolutional neural networks is that the input to the network is the raw data, or images in this case, rather  than feature values designed by astronomers \citep{krizhevsky2012imagenet}. This enables the network to learn and generate a hierarchy of features with minimal information loss \citep{oquab2014learning}. Each layer of convolutions learn different features. For example, the first few layers learn simple feature such as edges and corners. Successive layers combines these elementary features into more complex features to generalize the input data. This succession of feature learning generates a hierarchy of features \citep{masci2011stacked}. 
Figure \ref{cnn} shows a single layer of convolution in a CNN. 

CNNs are generally characterized by a third dimension in the network, often called the depth. In the case of image data, different kernels/filters can be learned at once in a given layer as shown in Figure \ref{cnn}. This enables learning of different features of the input data. Each learned kernel adds to the depth dimension of a layer. The general notion of CNNs is that the depth increases in the forward direction. The complexity of the features learned in each layer increases in the forward direction and finally they are combined into a fully connected layer.The final layer usually comprises of a cross-entropy function \citep{hagenauer1996iterative,de2005tutorial} for calculating the loss and a scoring layer at the end. The cross-entropy function is a decoding scheme used in information theory which is based on probability distribution of the sample classes. The final feature layer in a  neural network need to be decoded or converted to give the correct class of the training/ test sample. The cross-entropy does this job by looking at the distribution of the feature values. In this study we will be using a binary cross-entropy function since the models individually will be doing a binary classification.

Deep neural networks are neural networks which has many hidden layers, generally more than two \citep{hinton2012deep}. A convolutional neural network which have multiple layers of convolutions is termed Deep Convolutional Neural Network (DCNN) \citep{krizhevsky2012imagenet}. DCNNs have been widely used for image recognition and speech signal processing applications, and have been performing with exceptional accuracies \citep{krizhevsky2012imagenet,hinton2012deep,lecun2015deep}. In this study we have made use of DCNN for radio galaxy classification. 

Another interesting property of DCNNs is transfer learning \citep{yosinski2014transferable}. Transfer learning enables to train a network for a new application with few training examples. In areas like astronomy, it is often difficult to have clean, hand labeled datasets for different applications. It is possible to exploit the property of transfer learning in DCNNs to train a pre-existing model for a different classification problem. In addition to this, it also to improve the existing model without having to retrain from scratch, as opposed to other, shallow, machine learning methods. One of the main objectives of this study is to also provide a DCNN model that can be used for future transfer learning applications.

DCNNs have been used for different applications in optical astronomy such as star-galaxy classification \citep{kim2017star} and redshift estimation \citep{hoyle2016measuring}. In  recent work by \citet{dieleman2015rotation} a rotational invariant convolutional neural network was used for optical galaxy classification which gave near-human accuracy. These results provide motivation for the application of such techniques to radio astronomy as well.

\section{Network Model}
\label{network_model}
\begin{figure}[!htb]
\includegraphics[width=\textwidth]{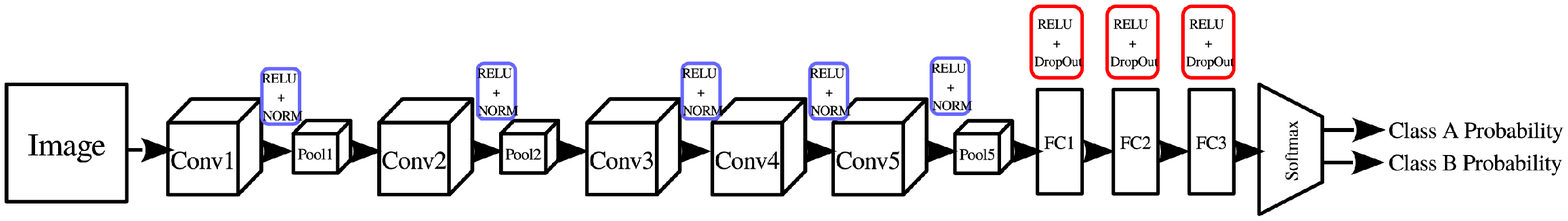}
\caption{Convolutional neural network architecture used in this study. The model takes in a single channel gray scale image of size 150 $\times$ 150 and outputs the classification score for two classes. The network has a total of 12 layers with 5 convolutional layers, 3 pooling, 3 fully connected and a scoring layer. The scoring layer produces the probability scores for two classes.}
 \label{alexnet}
\end{figure}

Neural network model design in general is considered to be a hyperparameter optimization problem. This simply means that there is no strict guideline for the design of a neural network. There is no rule to decide the number of hidden layers or number of neurons for a model. The model design is usually done with respect to the complexity of the data that is investigated. In the case of convolutional neural networks, model complexity is generally found to increase with complexity of the objects for classification. Even though simplified models can deliver good accuracies, their prospect for transfer learning \citep{oquab2014learning} are limited. This is because simple models generally have fewer layers of convolutions and activations. For this study we initially explored different simple models with up to 5 layers comprising of 3 - 4 convolutional layers. During training these models performed poorly with accuracies below 60\% which only is slightly better than random guess. One of the objectives of this study is to provide a model complex enough and standard that it can be used for studying more complex source morphologies with fewer training samples enabled by transfer learning. Therefore we chose to use a standard model which has been successfully used for different transfer learning models \citep{oquab2014learning}. We have used a slightly modified version of the Alexnet convolutional neural network \citep{krizhevsky2012imagenet}, see Figure \ref{alexnet}. This model has been successfully used for different image classification problems and it gave promising results (accuracies greater than 80 \%) for the initials tests we did. The advantage of this model is that it can be easily adapted to new classification problems and also can handle background noise in images \citep{joshi2012scalable,sukhbaatar2014training}. The original network is designed to work on color images and in our case we have modified it to work on single channel images. We have also made corrections to handle the image size and number of classes.

\begin{figure}[!htb]
\centering \includegraphics[scale=.5]{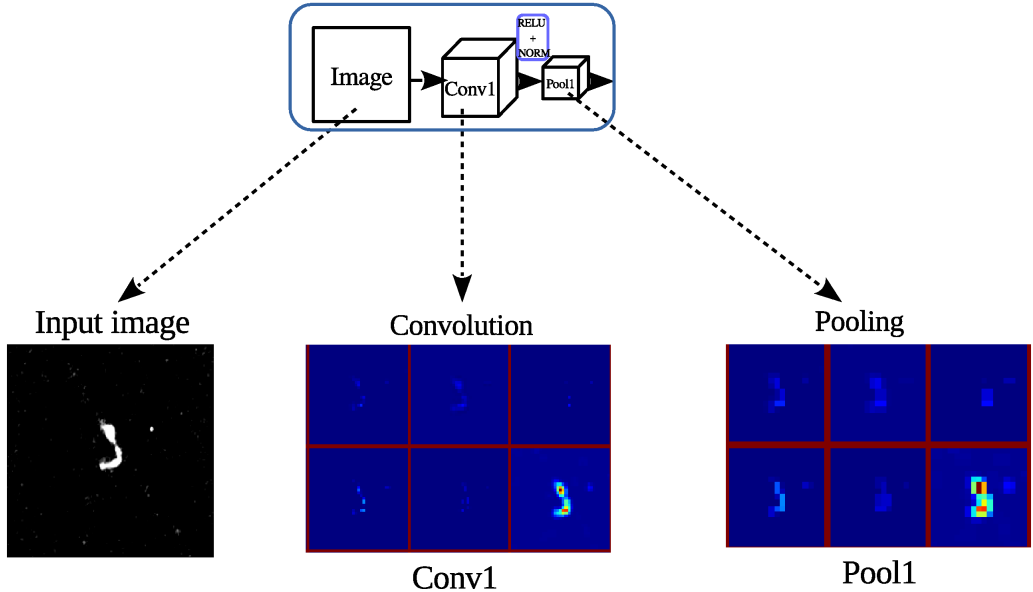}
\caption{Generalized illustration of convolution layer and pooling layer of the model. Convolution and pooling layers are the two main components of the DCNN used in this study. The figure shows the output of two such layers in the model. The left most column shows a preprocessed input image which us fed into the network. The second column shows just six filter outputs of the first convolutional layer. It can be seen from the highlighted pixels that each filter/ feature map has learned different features of the input image. Not all the filters have learned proper features which can be seen from the low values pixels in the output. The left most column shows outputs of the pooling layer of the corresponding convolutional layers. It can be seen that the pooling layer has down-sampled the output of the first convolutional layer. Similar operations happen in the rest of the network}
\label{layerexp}
\end{figure}

The model consists of 5 convolutional layers with three sets of maxpooling layers. The maxpooling layers are basically sub-sampling layers. This layer performs down-sampling of an input layer in a non-linear fashion so as to reduce computational complexity \citep{boureau2010theoretical} during forward propagation. Each convolution is followed by a Rectified Linear Unit (ReLU) \citep{nair2010rectified} which a kind of activation layer. This is  then followed by a normalization layer (NORM). The final pooled output (Pool5) is then passed onto a series of three fully connected layers which also have ReLU activations. Figure \ref{layerexp} shows what happens in the main component layers of the network mainly the convlutional layer and pooling layer.

Within the fully connected layers, neurons having weak connections are dropped out during training. Those neurons/nodes whose weights have a very small value  do not interact with other nodes. Such node connections are insensitive to weight updates and are termed as weak connections. These nodes can be discarded off the network since they do not influence the forward pass. This procedure is called \textit{dropout} and is a mechanism  to avoid over-fitting \citep{srivastava2014dropout}. 50\% dropout is carried out in this network, which is essentially removing 50\% weak neuron connections. 

The fully connected layer FC3 has a depth of two in contrast to the other fully connected layers. The final layer during training  calculates the cross entropy loss \citep{gold1996softmax} and during validation outputs a Softmax probability score for the two classes. Thus the network has 12 different layers including the different convolutional layers, pooling, fully connected and softmax layer. The functional description of each layer, kernel sizes and learned parameters are given in Table \ref{table1}.

\begin{table}[!ht]
\centering
\caption{Table of Layer parameters and functions}
\label{table1}
\begin{tabular}{ccccc}
\hline
\textbf{Layer name} & \textbf{Function}     & \textbf{Depth} & \textbf{Kernel size} & \textbf{Parameters}      \\ \hline \hline
Conv1               & Convolution           & 96             & 11x11                & \multirow{2}{*}{11712}  \\ \cline{1-4}
Pool1               & Max Pool              & 96             & 3x3                  &                          \\ \hline
Conv2               & Convolution           & 256            & 5x5                  & \multirow{2}{*}{307456} \\ \cline{1-4}
Pool2               & Max Pool              & 256            & 3x3                  &                          \\ \hline
Conv3               & Convolution           & 384            & 3x3                  & 307456                  \\ \hline
Conv4               & Convolution           & 384            & 3x3                  & 663936                  \\ \hline
Conv5               & Convolution           & 256            & 3x3                  & \multirow{2}{*}{442624} \\ \cline{1-4}
Pool5               & Max Pool              & 256            & 3x3                  &                          \\ \hline
FC1                 & Fully Connected Layer & 4096           & \multirow{3}{*}{}    & 16781312               \\ \cline{1-3} \cline{5-5} 
FC2                 & Fully Connected Layer & 4096           &                      & 16781312               \\ \cline{1-3} \cline{5-5} 
FC3                 & Fully Connected Layer & 4096 x 2       &                      & 8194                    \\ \hline
Softmax             & Softmax Layer         & 2              & \multicolumn{2}{c}{}                            \\ \hline \hline
\multicolumn{4}{c}{\textbf{Total Number of parameters learned}}                              & 35881666               \\ \hline
\end{tabular}
\end{table}

In Table \ref{table1}, we can see that as the number of convolutional layers increases in the forward direction, and also that the total number of parameters the network has to learn increases dramatically. This is one drawback of deep neural networks : one needs large amount of memory to hold these parameters during training. At present, this challenge has been overcome with the advent of GPUs which can do fast computation of error gradients in a neural network while also having enough memory to hold large number of parameters during the training phase. During the testing phase, where there is only forward computation and no backward computation, the memory issue is reduced. Memory overflow can only happen when the batch size of the input is too large or the image size is too big compared to the total available memory. 

\subsection{Pre-processing Sample Images}
\label{preprocessing_sample_images}
\begin{figure}[!htb] 
\includegraphics[scale =.65]{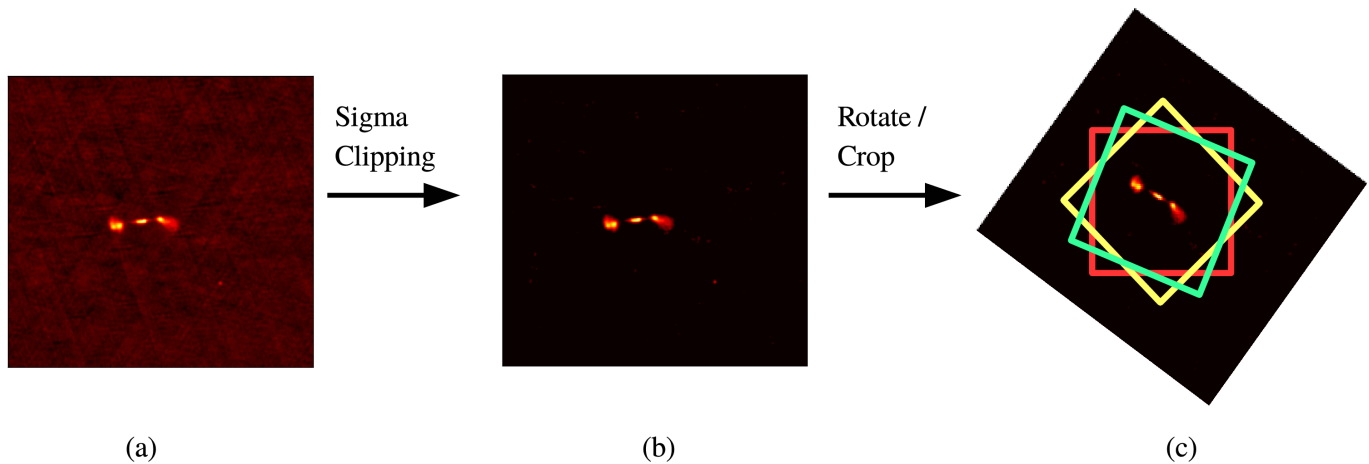}
\caption{The different stages of image pre-processing  before training / testing the network. (a) is the image of the object which is of size 300x300 pixels. (b) The pixels below some noise threshold are suppressed with sigma-clipped statistics. (c) The image is cropped from the center to size 150x150 pixels and similarly for the rotated and flipped versions.The smaller boxes show different cut outs at different rotations.}
\label{preprocess}
\end{figure}

Certain image preprocessing steps are desirable before the image is fed into a convolutional neural network or any other machine learning algorithms. This is usually done for maintaining the homogeneity of the sample space. This procedure has specific importance with convolutional neural networks because the neurons behave like visual receptors similar to human visual receptors. The basic idea is that if a human is able to \textit{see} an object, the network should also be able to \textit{see} it. The different stages are shown in Figure \ref{preprocess}.

First,  the sigma-clipped statistics\footnote{Using the Astropy functionality for sigma clipping,$  http://docs.astropy.org/en/stable/api/astropy.stats.sigma\_clipped\_stats.html\\ \#astropy.stats.sigma\_clipped\_stats$} of each image are estimated in order to calculate the background noise and flux levels. With sigma-clipped statistics, pixels above certain sigma  level from the median are discarded or nulled. In this study all values below a 3$\sigma$ level of the background were cut-off by suppressing those values to zero so as to highlight the contribution of the source and remove any unwanted background noise.The value for sigma-clipping was chosen by training and testing the model at different sigma values. After different iterations and tests we found that the value of 3-sigma was better than 2-sigma or 5-sigma. In all cases other than 3-sigma the model had an accuracy less than 60\%.

One of the important requirements  of neural networks is a large number of training samples. With  less than 500 training samples in total, it is impossible for the network to learn the features  and generalize the different classes. To overcome this issue, training samples can be over-sampled while preserving the labels by rotating and flipping each sample \citep{krizhevsky2012imagenet}. Generally speaking to train machine learning algorithms, the nominal number of training examples is in order of 10000. For deep neural networks this number is much larger. In this particular case where we have only few training samples the label preserving oversampling will help create a large training set to train the network optimally. Here each sample which is either a rotated or flipped version of a specific image is treated as a unique training sample. This procedure also helps the network learn rotational invariant features of the samples in the convolutional layers. Another issue that need to be addressed along these lines is the class imbalance in the synthetic training set. Machine learning algorithms require fairly equal number of training samples for each class for the model to have good balance between bias and over-fitting \citep{duda2012pattern}. The model will tend to over-fit if there is large number of training samples for a specific class compared to the other. In this study we have balanced the number of training samples by suitably choosing the number of angles for the rotations and their flipping. This procedure will help to have more evenly distributed samples in the parameter space learned by the convolutional layers. 

The different pre-processing steps, shown in Figure \ref{preprocess} can be summarized as follows. The downloaded images were 300x300 pixels in size. The images were rotated by small angles  in steps of either 1$^{\circ}$, 2$^{\circ}$ or 3$^{\circ}$ such that number of samples for all the three classes were roughly equal. The number of rotation steps were chosen such that all the different classes had roughly equal number of bootstrapped training samples so as to minimize any over-fitting issues in the model. Afterwards a 150x150 patch centered on the source was cut out from the main image. Flipped and rotated versions of the samples were also generated. 

\subsection{Training}
\label{training}
To evaluate machine learning models, the whole data is generally split into two parts. The first part of the data is used to train the machine learning model. The second part of the split data is used to validate the performance of the trained model. This part of the data is known as validation or test data. It is a general practice to take larger portion of the data to be used for training and the remaining for validation. None of the samples in the validation set are seen by the model during training.
In this study the complete dataset for the 3 classes was split with a 70-30 ratio, where 70\% of the original data were taken for training and 30\% for validation. So the actual number of training samples were 125 FRIs, 227 FRIIs and 177 Bent-tailed. The number of validation samples were 53, 57 and 77 for FRI, FRII and Bent-tailed radio galaxies respectively. The data oversampling and augmentation were done for the training set. Thus, the number of samples for FRIs were 45000, 40680 FRIIs and 31860 Bent-tailed `sources'.  Afterwards the  training data was again split randomly for training and testing with a 80-20 ratio. Therefore the number of samples that went into training from this second split were 36000 FRIs, 32688 FRIIs and 25488 Bent-Tailed `sources'. The training and test samples, being over-sampled versions of the same training images, have overlap since they were generated from the same base samples, but the validation samples which were split from the original 70-30 split were never seen by the network during training.

The network was implemented with the deep learning package called Caffe \citep{jia2014caffe} which is widely used in computer vision applications. We used an NVIDIA forked version of Caffe to support training on multiple graphical processing unit (GPU) cards.  Images in Portable Network Graphics(PNG) format were converted to Lightning Memory-Mapped Database (LMDB) for quick access to memory. The training was done on a machine with an Intel(R) Xeon(R) CPU, 260 GB memory and  four TITAN-Black GPUs with 12GB RAM each.

\begin{figure*}[ht!]
\centering
\gridline{\fig{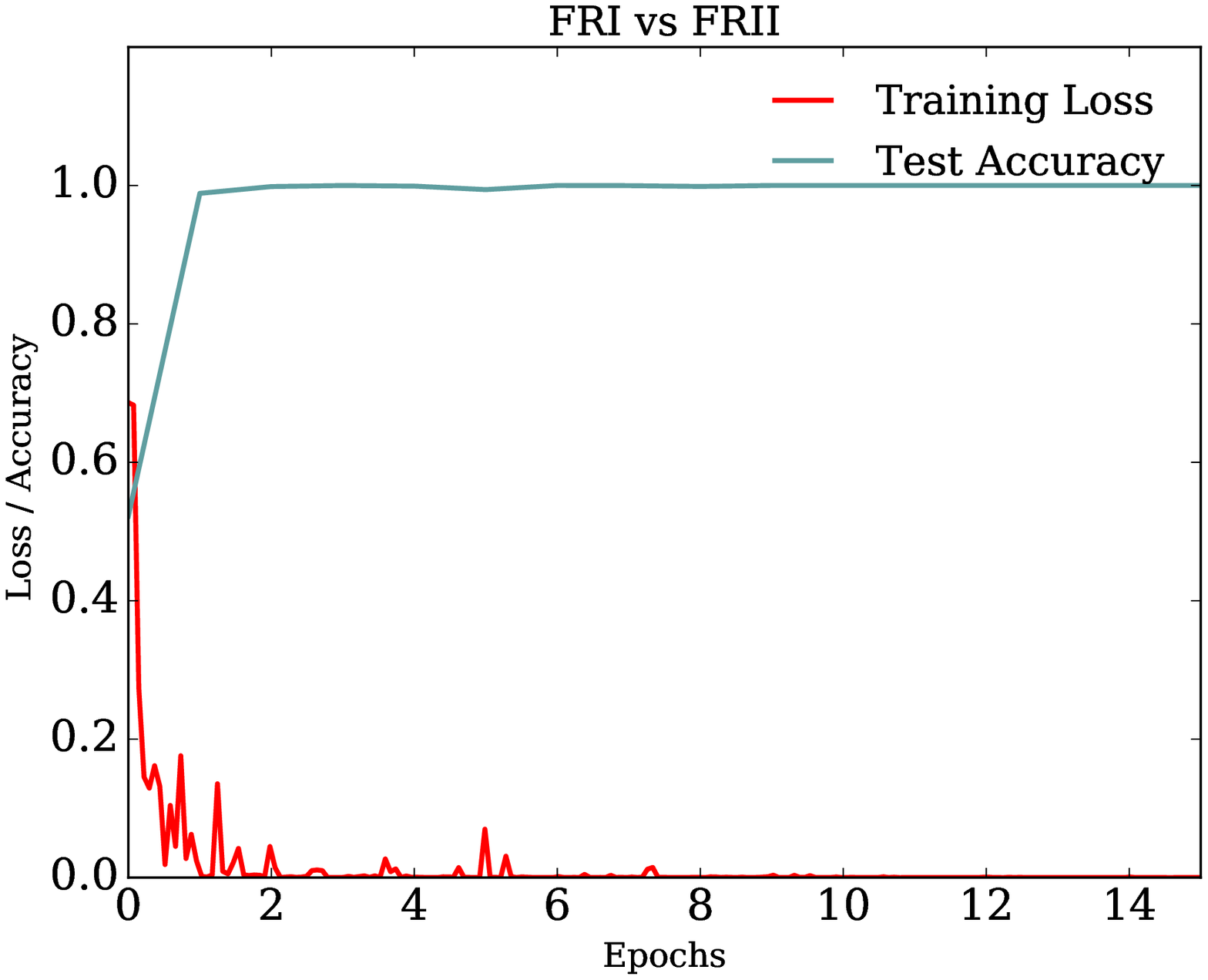}{0.35\textwidth}{(a)}
\fig{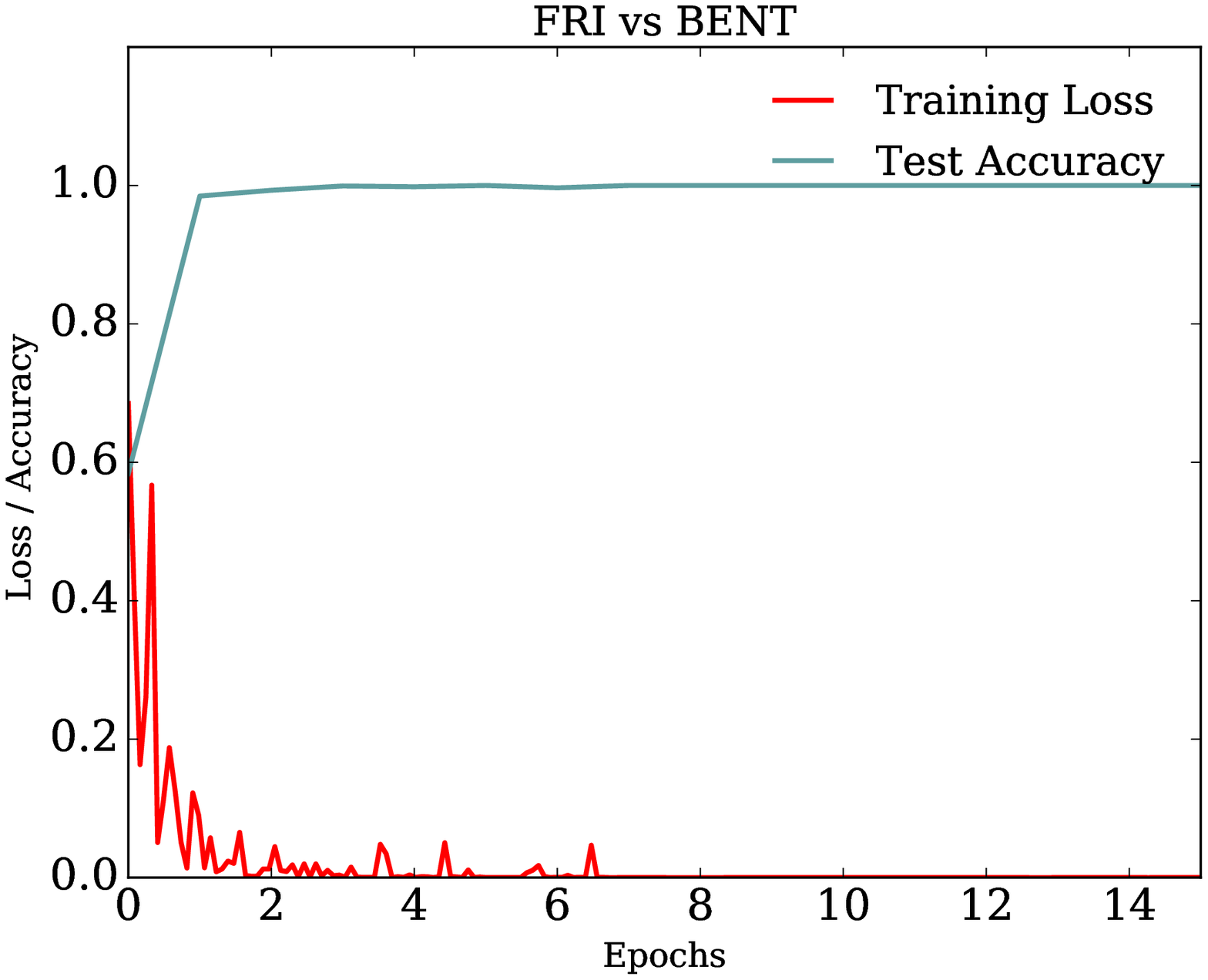}{0.35\textwidth}{(b)}
\fig{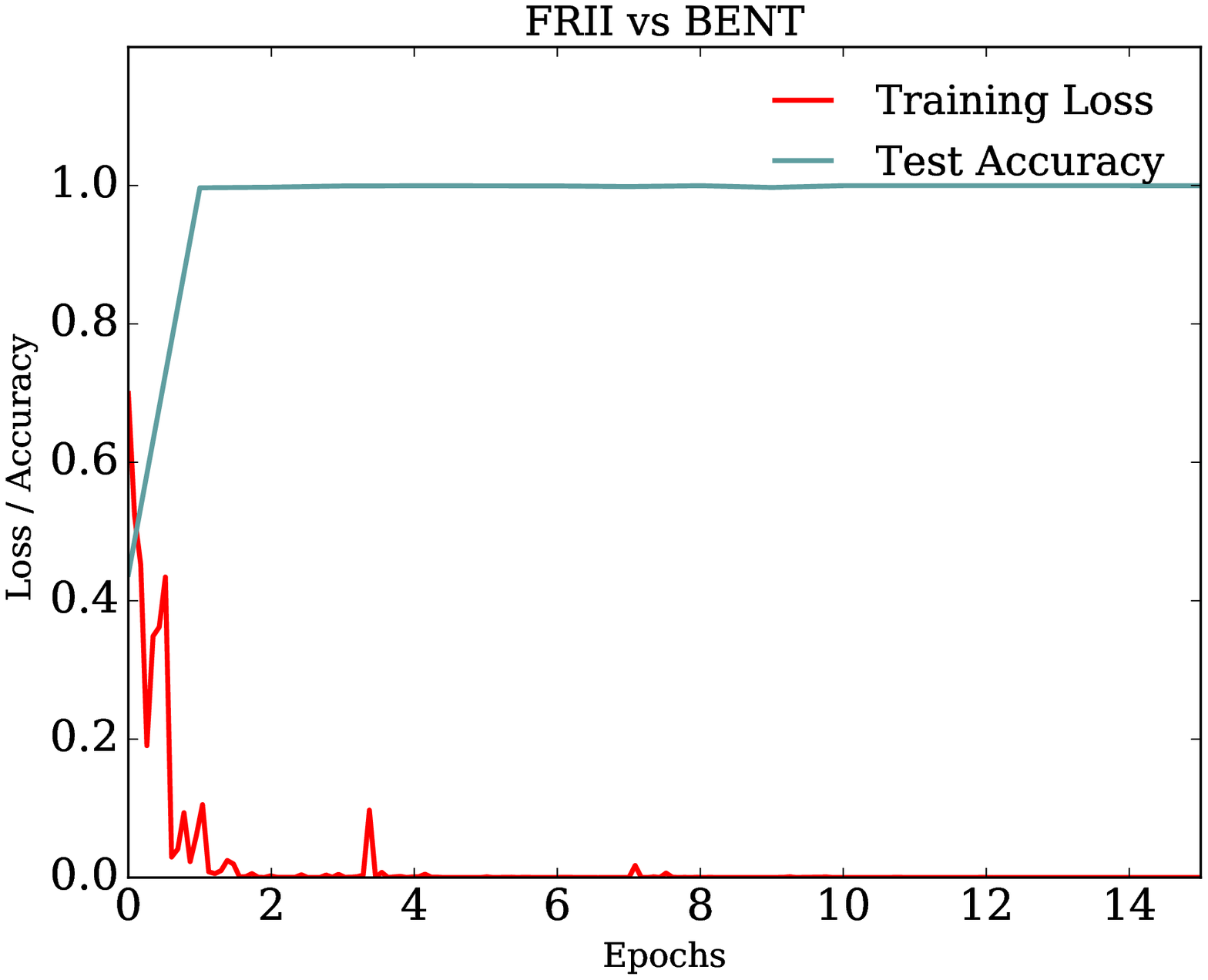}{0.35\textwidth}{(c)}
}
\caption{Learning curves showing the training loss and test accuracy for the three different binary classification models. It can be seen that the test accuracy and training loss saturates after 10 epochs for all the three models.(a) shows the training and testing accuracy for FRI vs FRII classification. Similarly (b) and (c) shows the learning curves for FRI vs bent-tailed and FRII vs bent-tailed classifications, respectively.}
\label{learningcurve}
 \end{figure*}

The kernels of each layer were initialized with random Gaussian values. We used a stochastic gradient descent algorithm \citep{duda2012pattern} with a batch size of 100 for training. The batch size determines the number of samples that is used for a single forward pass before calculating the backpropagation error by the stochastic gradient descent algorithm. The best learning rate was a step function with base learning rate of 0.01. The training was done for 30 epochs and a validation of the network was done during every epoch to keep track of the learning performance. The learning curves for the three binary classification models are shown in Figure \ref{learningcurve}.

The learning curves give a measure of the performance of the machine learning model for the training and testing data \citep{perlich2011learning}. The training loss, which is a negative log-likelihood, is calculated from the cross entropy error \citep{hinton2006reducing} and is given as
\begin{equation} \label{loss-eqn}
L(w) = - \frac{1}{N} \sum_{n=1}^{N} \left[y_{n} log  \hat{y_{n}} + (1 - y_{n}) log (1 -\hat{y_{n}}) \right]
\end{equation}

In equation \ref{loss-eqn}, $N$ is the number of training samples, $\mathit{w}$ is the weight vector, $\hat{y_{n}}$ the expected output and $y_{n}$ is output during a forward pass. The stochastic gradient algorithm minimizes the error $L(w)$ by properly adjusting the values of the weight vector $\mathit{w}$. Thus the training loss gives us an idea of how well the model is learning over each iteration or epoch. 

The test accuracy is determined when, for each epoch, the model parameters are fixed and no learning takes place, while the model is tested against the test data. The test accuracy  computed for each epoch is given as 
\begin{equation}
Accuracy = \frac{1}{N} \sum_{n}^{N} \delta \left\lbrace \hat{l_{n}} = l_{n} \right\rbrace ,   \delta \left\lbrace condition \right\rbrace 
\begin{cases}
1 & \text{if condition} \\
0 & \text{otherwise}
\end{cases}
\end{equation}
$N$ being the number of test samples, $\hat{l_{n}}$ is the predicted class label for the $n^{th}$ sample and $l_{n}$ is the true label. 

With four GPUs, the training for each test cases took around 1.2 hours. In all the cases we observed that the accuracy tended to saturate to high values after 10 epochs and the training loss fell steeply to very low values. This is because each model is a simple binary classification problem and the network tends to learn quickly without too many training epochs.

\subsection{Filter Visualization}
\label{filter_visualization}
\begin{figure}[!ht]
\centering
\includegraphics[scale=.5]{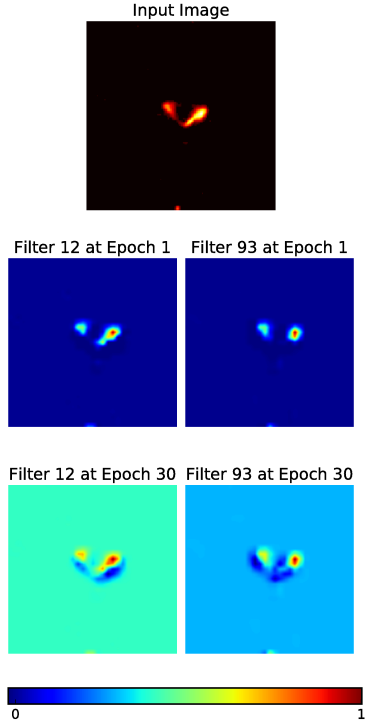}
\caption{Visualization of the output of two random filters from the first convolutional layer of the network. During the initial epochs the network, the weights had small values (middle row); and as the learning progressed the network learned to distinguish between the background and the object. It can be seen from the visualization in Epoch 30 (bottom row) that the weights had solidified with larger values for the object and the background. The filter values here are scaled from their actual values for visualization.}
\label{filter}
\end{figure}

Filter visualization of the network model helps to understand what happens as the learning progresses with each epoch. Different filters learn different properties/features of the object. Looking at the filter visualizations and the learning curves in Figure \ref{learningcurve}, we can see how the confidence of the network improved with each epoch. During the first few epochs, the network had confusion between the object and the background, and with further learning it gains confidence in distinguishing both. This is shown in Figure \ref{filter}. 

Figure \ref{filter} shows the output from two random filters in the first convolutional layer at the first epoch and the last epoch. It is evident that the network has learned to distinguish between the object and the background. Filter 12 has learned the sharp / high frequency features of the radio galaxy and Filter 93 more of the smooth features. In both cases the network has learned to recognize the radio galaxy in the final epoch. Initially only a part of the radio galaxy is recognize,d while later on the major parts of the galaxy are being recognized. The filter outputs in the initial layers are fairly easy to explain, but as one goes into further layers in the forward directions, the filter visualizations are more difficult to explain \citep{zeiler2014visualizing}.

\section{Classification Model}
\label{classification_model}
With three classes of objects in the study, we trained three different models for binary classification namely FRI vs FRII, FRI vs Bent-tailed radio galaxies and FRII vs Bent-tailed radio galaxies. Since the actual sample size of the three objects for training is highly imbalanced, there will be a general bias in the models having comparable and large sample numbers. Initially we trained a single DCNN to classify the three objects together. So the single model would predict if the given sample was either FRI or FRII or Bent-tailed radio galaxy. Even though the bootstrapping procedure that generated synthetic training samples to overcome the issues of few training examples and class imbalance problem, the single model we trained to classify the three classes performed inefficiently during training and validation. During training the model showed large training loss which a clear indication of poor learning and over-fitting. It was found during training that the training error increased during each epoch and the corresponding test accuracy in each epoch went below 60\%. This confirmed that the model was not performing well. 

The first solution to minimize this large training loss with a single model was change the loss function used for training. The Info-Gain loss function is designed and suitable for tackling class imbalance in convolutional neural networks \citep{jia2014caffe}. We experimented the info-gain loss function with different parameters and retrained the network. It was found that with all the different parameters we tried to optimize with this loss function, the network did not learn the task optimally and both the training and validation results were poor. We then broke the three class classification problem into three binary classifications which performed better in terms of individual classifications. 

\begin{figure}[!htb]
\centering
\includegraphics[scale=.4]{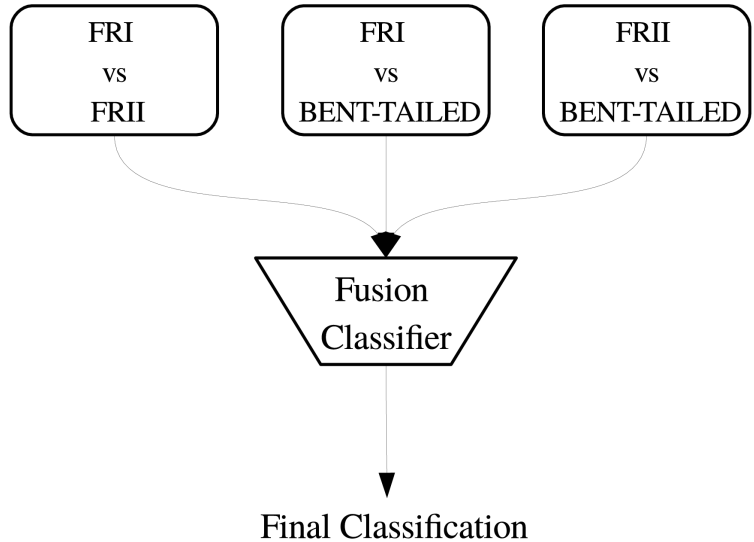}
\caption{Fusion model with majority voting ensemble classifier which combines the predictions from the three binary classifier models to make the final prediction. The figure shows the individual predictions from the three binary classifiers being fed into a fusion classifier which gives the final classification. This model is ideal in situations where there the individual models have a slight bias and also beneficial to identify odd inputs.}
\label{fusion}
\end{figure}

To overcome the issue of tuning model complexity, we have made use of a fusion classifier which is basically a majority voting classifier\citep{dietterich2000ensemble}. This is illustrated in Figure \ref{fusion}.

The fusion model takes the individual predictions of the binary classifiers and their corresponding probabilities to make the final prediction. In general, if for a given sample, two classifiers predict the same class with high probability then the final class will be the same. But if the three binary predictions are different and have mixed or low probabilities for their predictions, such samples will be rejected and classified as ``\textit{strange}" objects, and their probability value will be set to zero. This allows users to find objects of potentially interesting or confusing morphology. 
\pagebreak
\section{Results \& Discussion}
\label{results_and_disctussion}
The performance of the fusion model is evaluated on the basis of the classification precision, recall and $F_{\beta}$ score in percentage. The precision gives a measure of correctly classified samples and is given as,
\begin{displaymath}
\mathrm{Precision} = \frac{TP}{TP + FP}
\end{displaymath}
$TP$, the true positives is the number of correctly classified test samples and $FP$, the false positives, is the number of incorrectly classified test samples. The recall which is also called the sensitivity of the classifier is given as,
\begin{displaymath}
\mathrm{Recall} = \frac{TP}{TP+FN}
\end{displaymath}
where FN is the number of false negatives in the prediction. The recall value can be used to check if the model is over-fitting. For a good model, the precision and recall should be high. The $F_{\beta}$ score is a measure that combines both values of precision and recall. The $F_{\beta}$ score is expressed as
\begin{displaymath}
F_{\beta} = (1+\beta^{2}) \cdot \frac{Precision \times Recall }{Precision + Recall}
\end{displaymath}
In our test cases we make use of $F_{1}$ score where $\beta = 1$. For a good classification the $F_{1}$ score is close to 100\%. 

\begin{figure*}[!htb]
\centering
\includegraphics[scale=1]{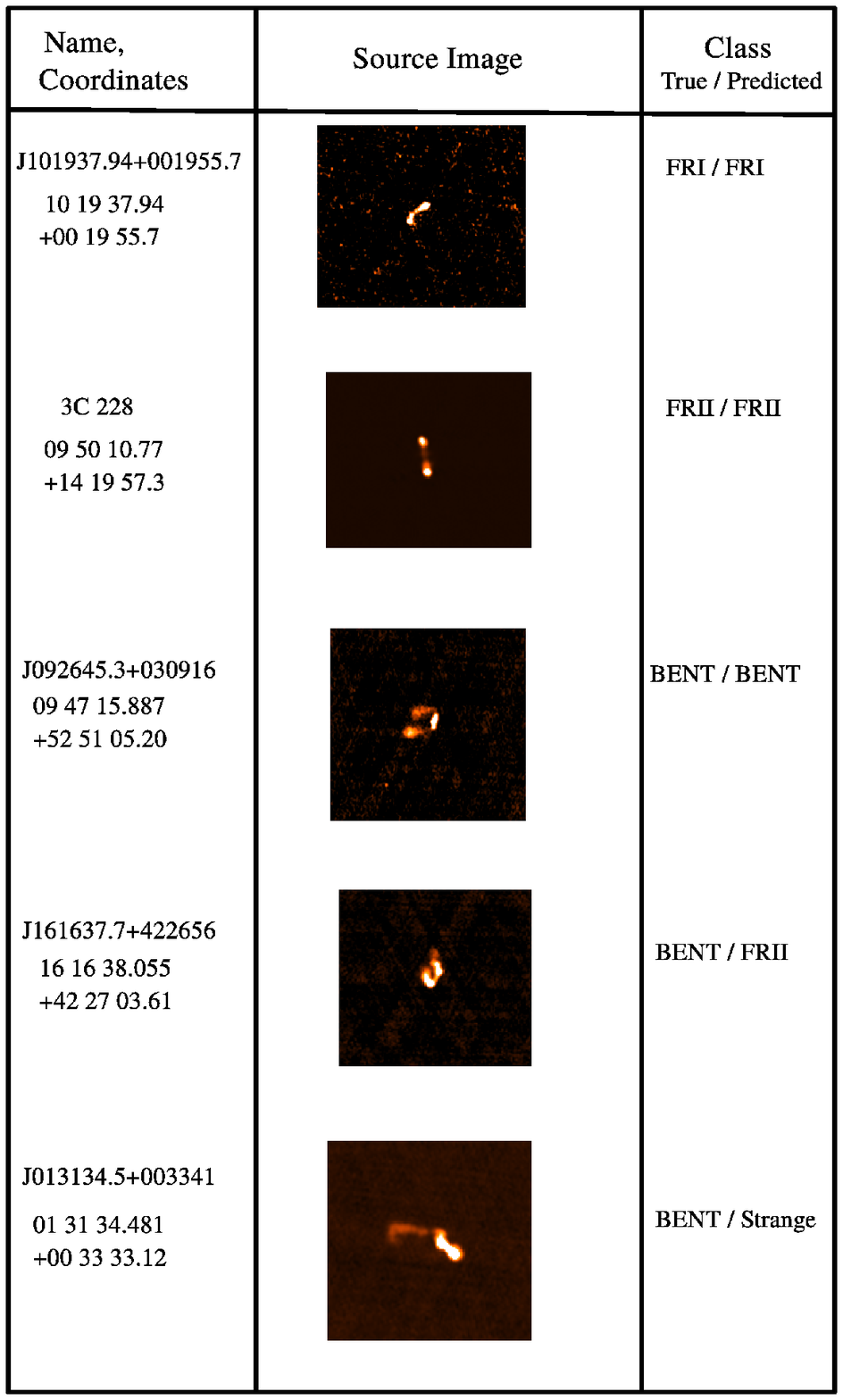}
\caption{Sample predictions made by the classifier model. The first column shows the name of the object and their coordinates, the middle column the image cut out and the left column shows the true class and the predicted class.}
\label{sampleres}
\end{figure*}

The trained model was used to classify 30\%  validation samples from the FIRST dataset using the fusion model. Table \ref{results}  shows the classification results for the FIRST samples in the validation set.  The ``support" column shows the number of test samples in each class. Figure \ref{sampleres} shows some of the predictions by the classification model. The average score for precision, recall and F1 score are calculated as a weighted average from the receiver operating characteristic (ROC) \citep{bradley1997use} of the predictions.

From Table \ref{results}, we can see that for all validation samples the models show excellent precision, recall and F1 score. The average precision is 88\% and average recall is 86\%, with an F1 score of 86\%.  The results of the fusion classifier can be understood as follows. To be assigned a class, a source needs to be identified as belonging to that class in both the individual classifications in which that particular class features. 

The bent-tailed radio galaxy classification shows a very high precision at 95\%, meaning that most of the classifications labeled as bent-tailed have been identified correctly. The recall for the bent-tailed class is poorer at 79\% , which implies that the algorithm was not able to identify all bent-tailed radio galaxies in the validation sample. 

The FRI radio galaxy classification shows both high precision and recall - this implies that the network model is able to identify FRI radio galaxies without much confusion.

\begin{figure}
\gridline{\fig{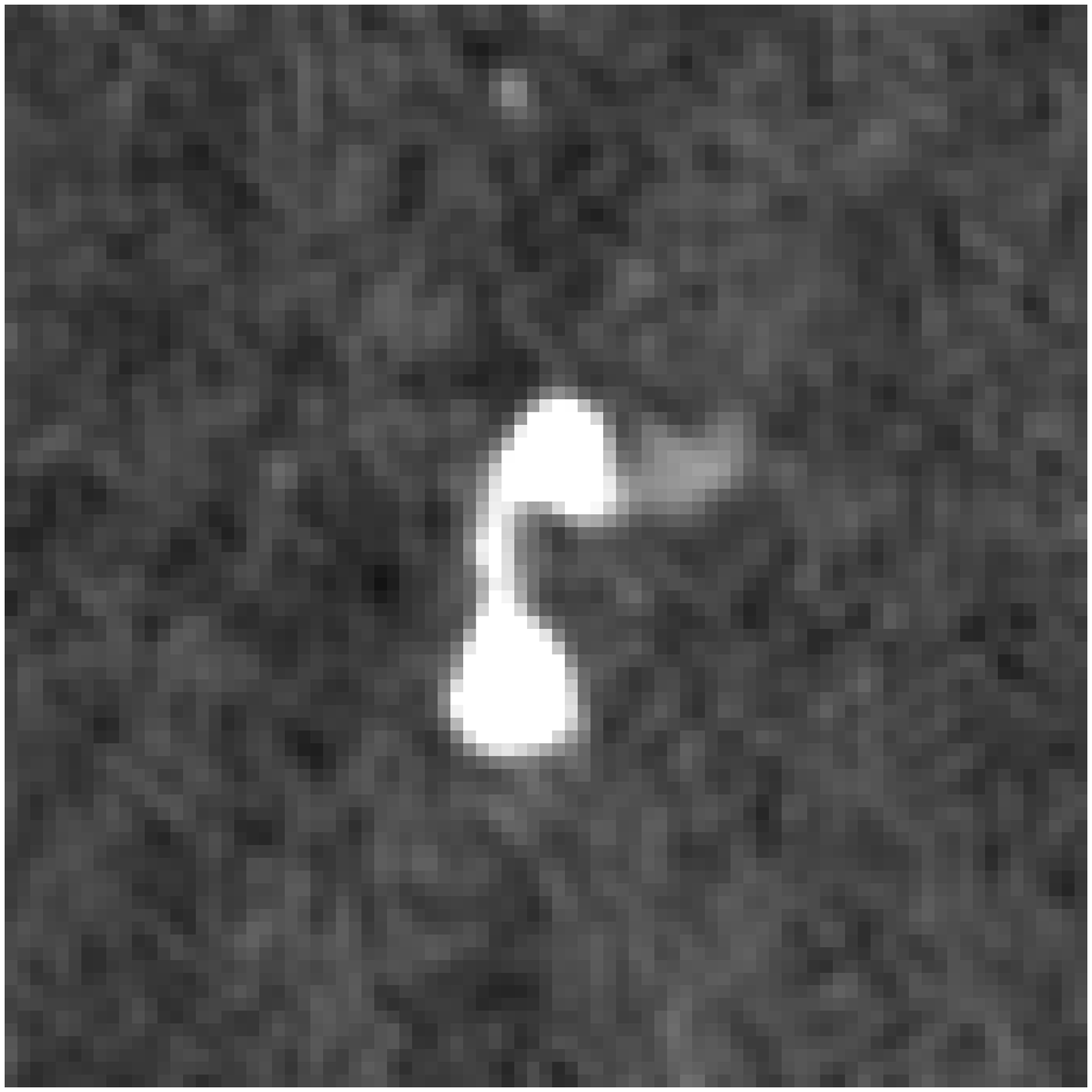}{0.2\textwidth}{(a)}
                   \fig{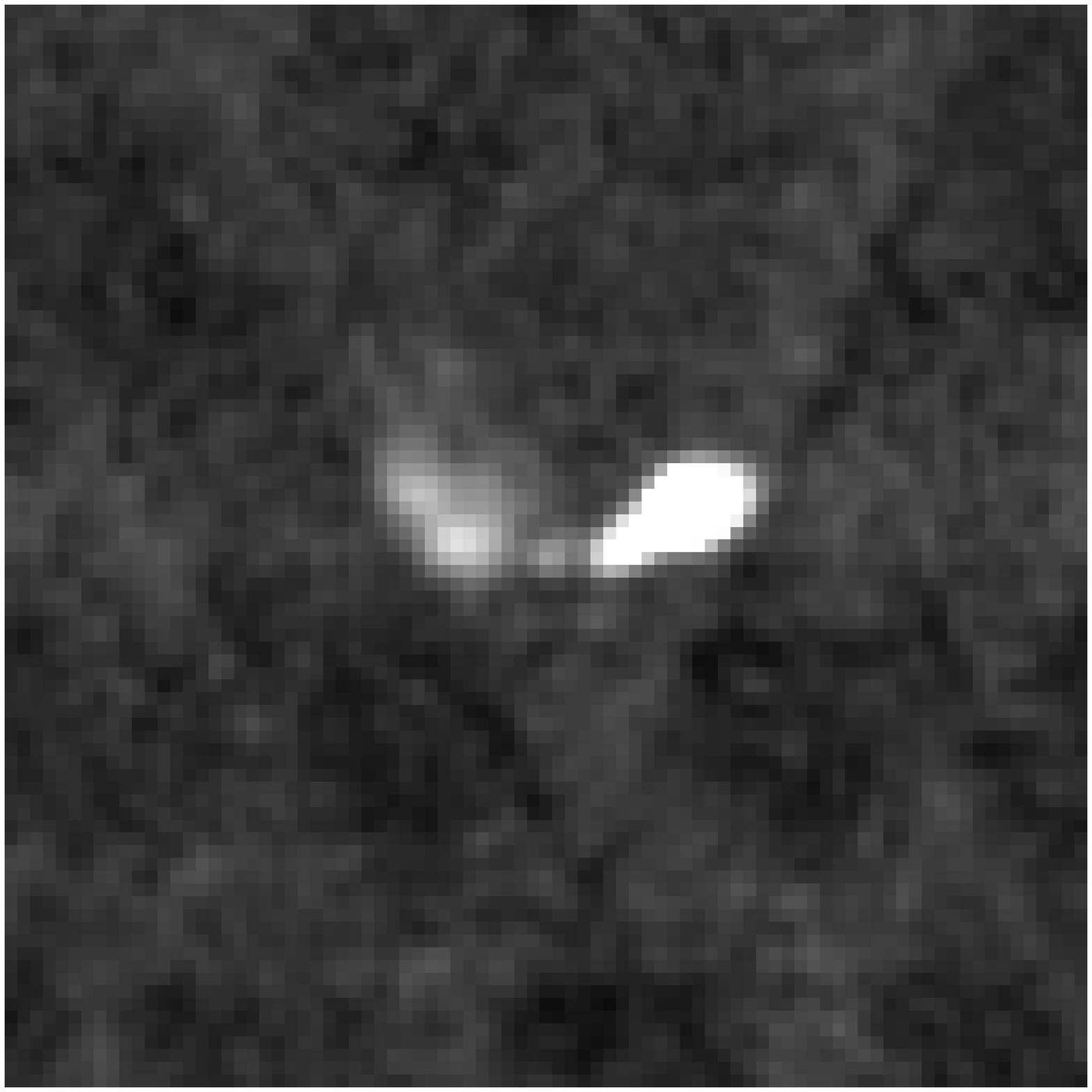}{0.2\textwidth}{(b)}
                   \fig{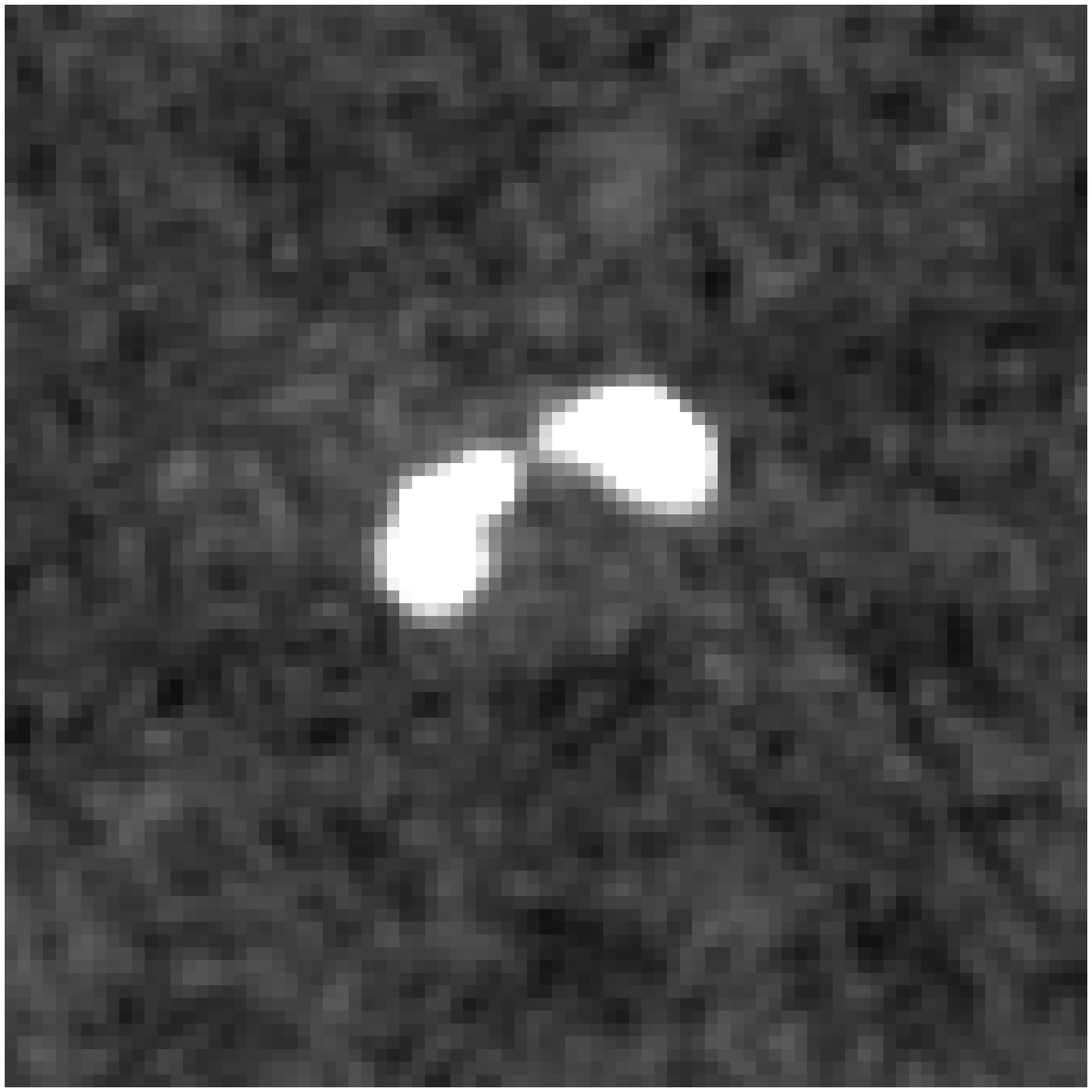}{0.2\textwidth} {(c)}
                   \fig{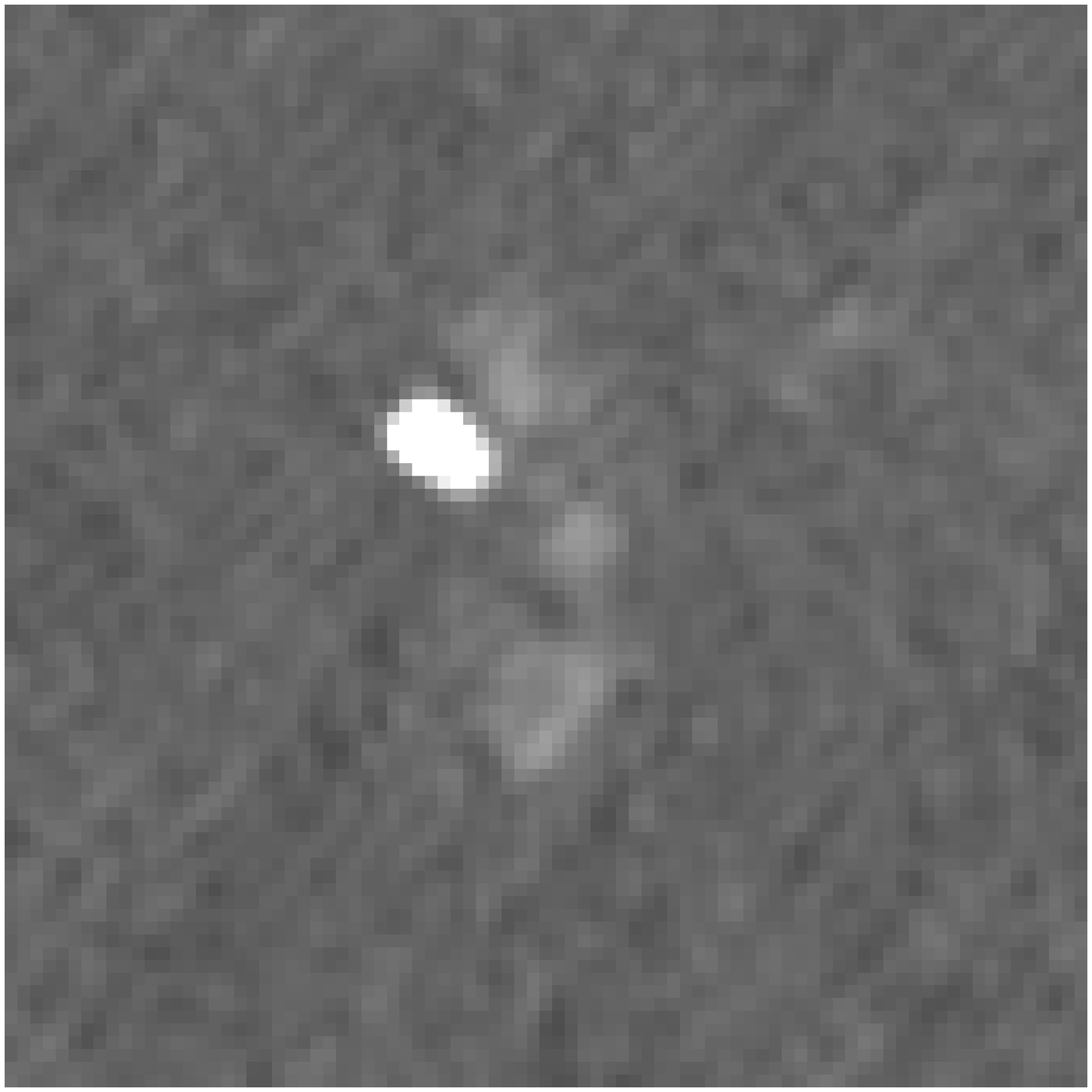}{0.2\textwidth}{(d)}}
\gridline{\fig{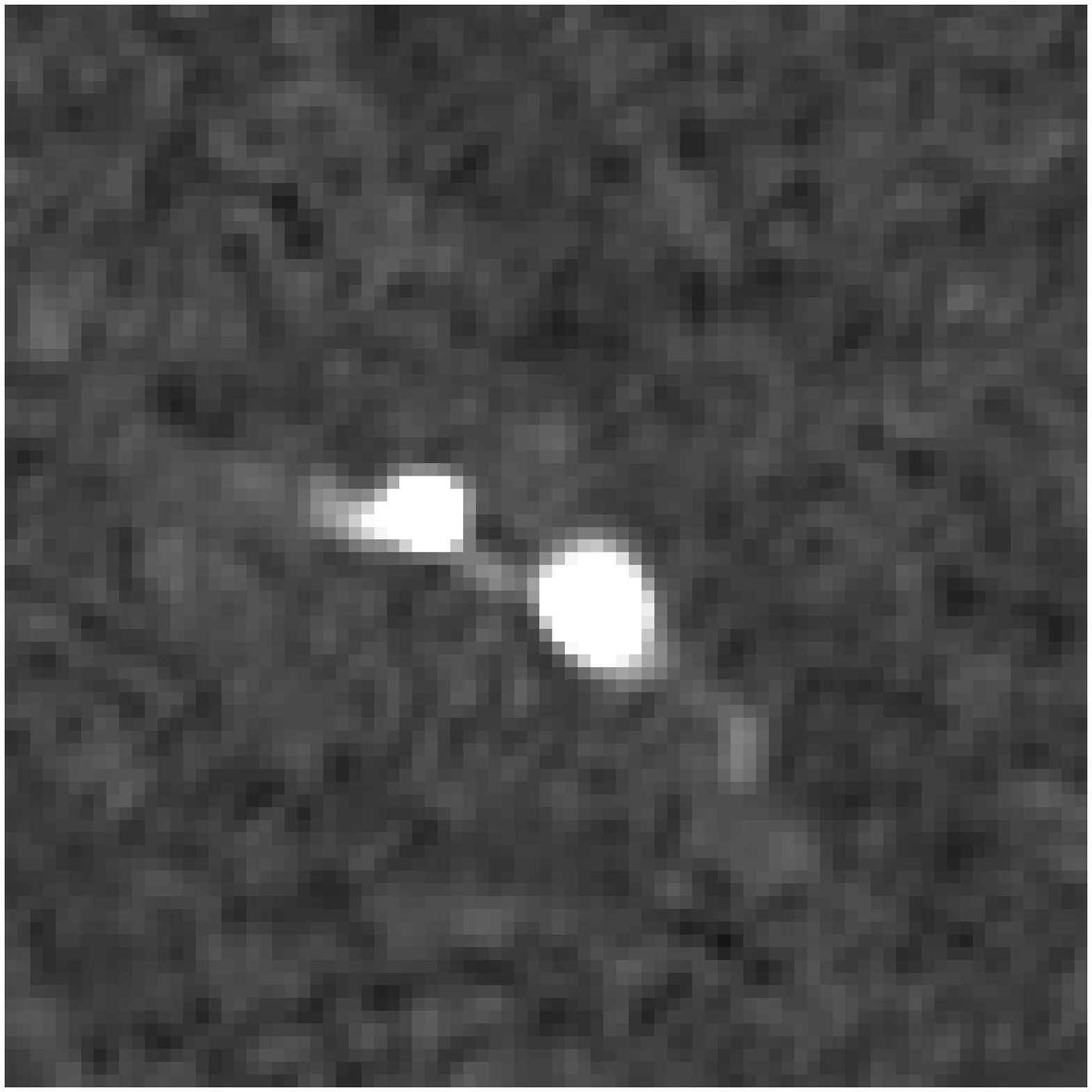}{0.2\textwidth}{(e)}
                   \fig{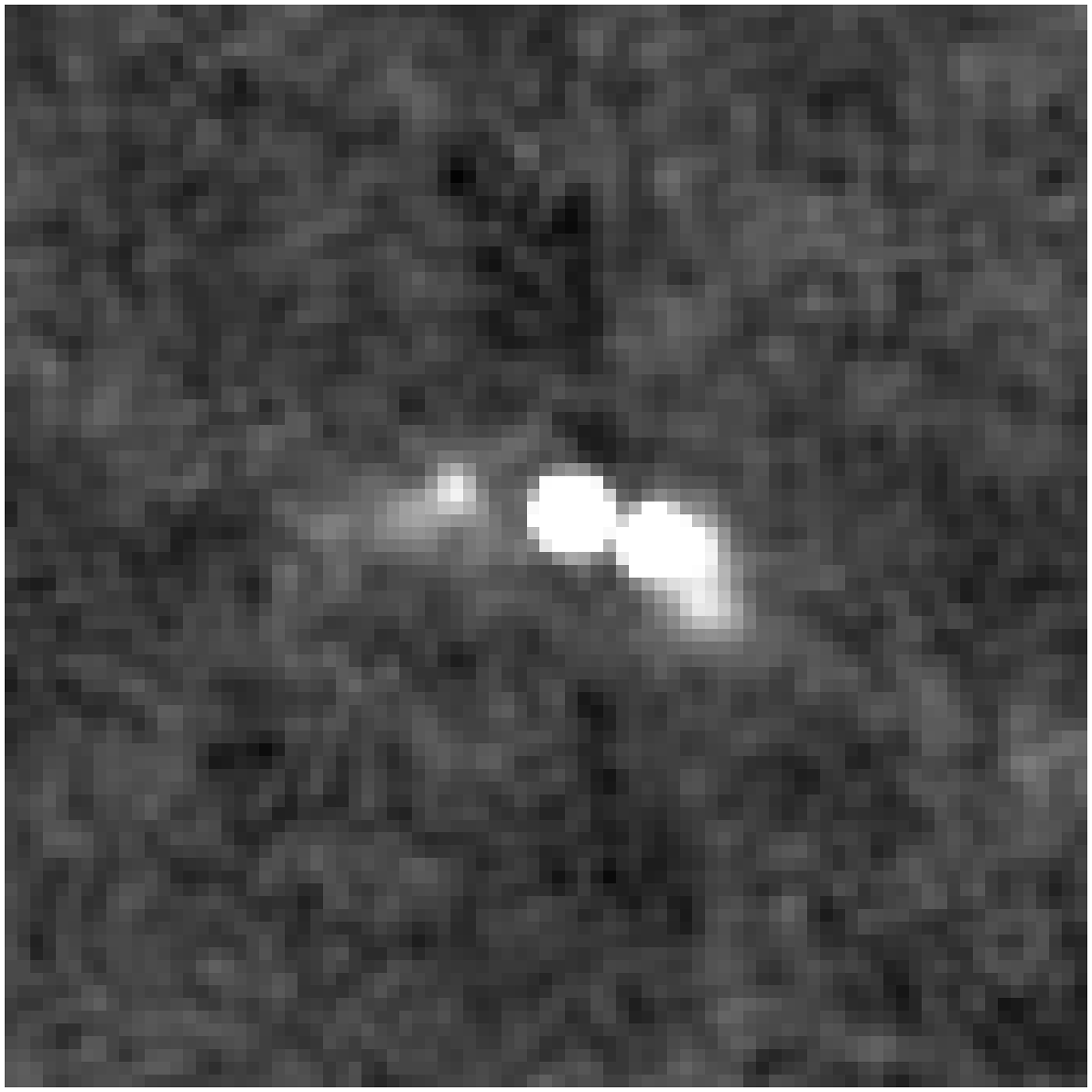}{0.2\textwidth}{(f)}
                   \fig{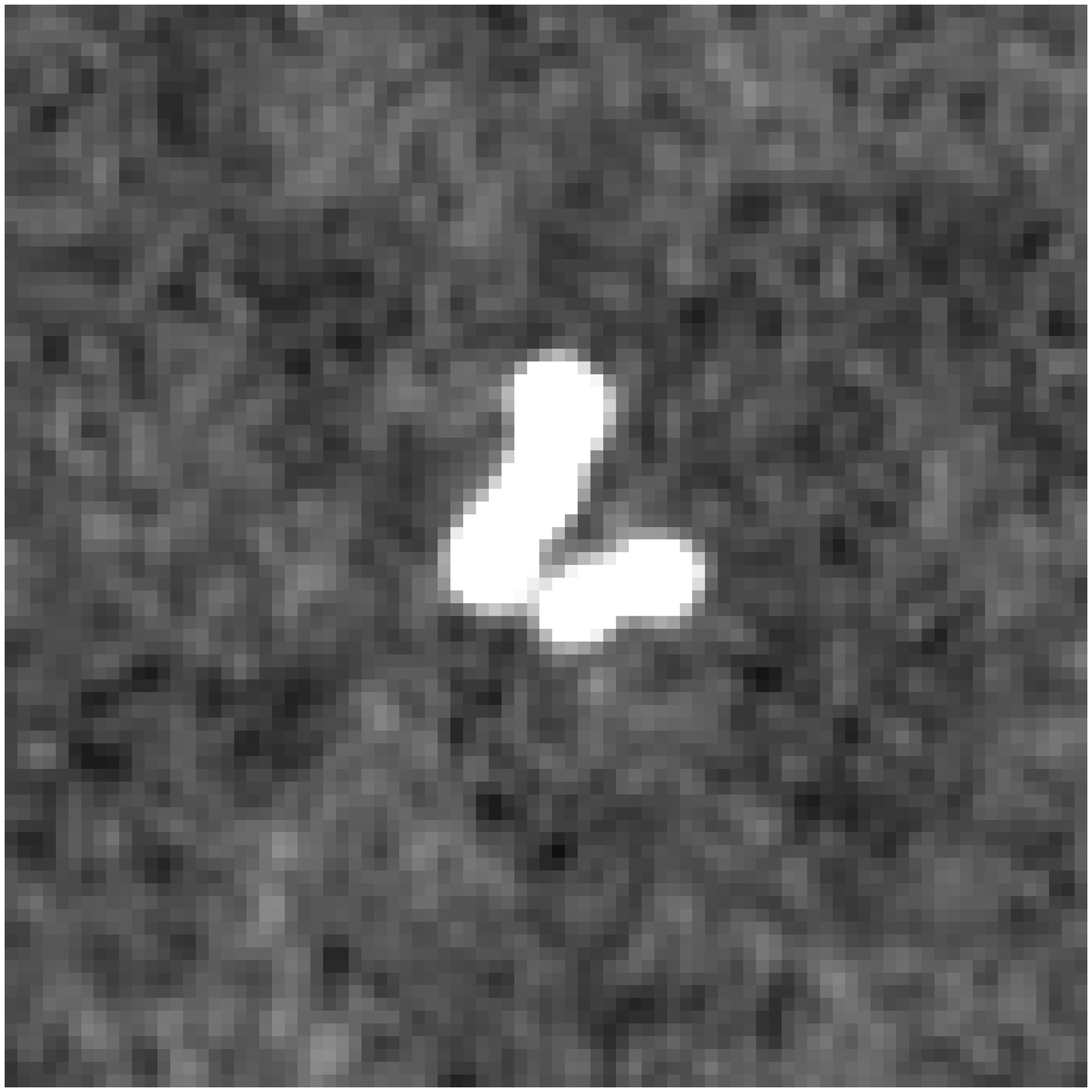}{0.2\textwidth}{(g)}
                   \fig{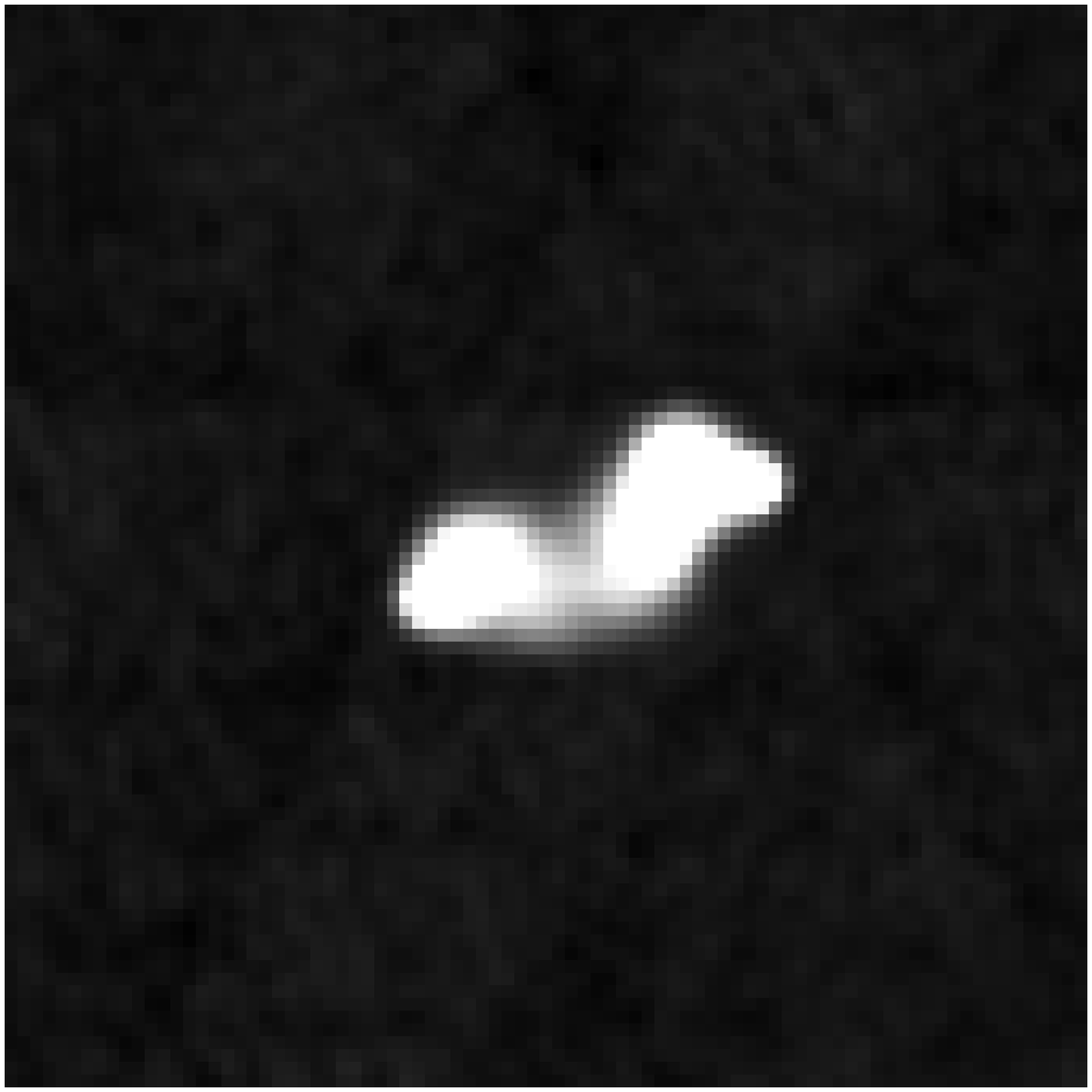}{0.2\textwidth}{(h)}}

\gridline{\fig{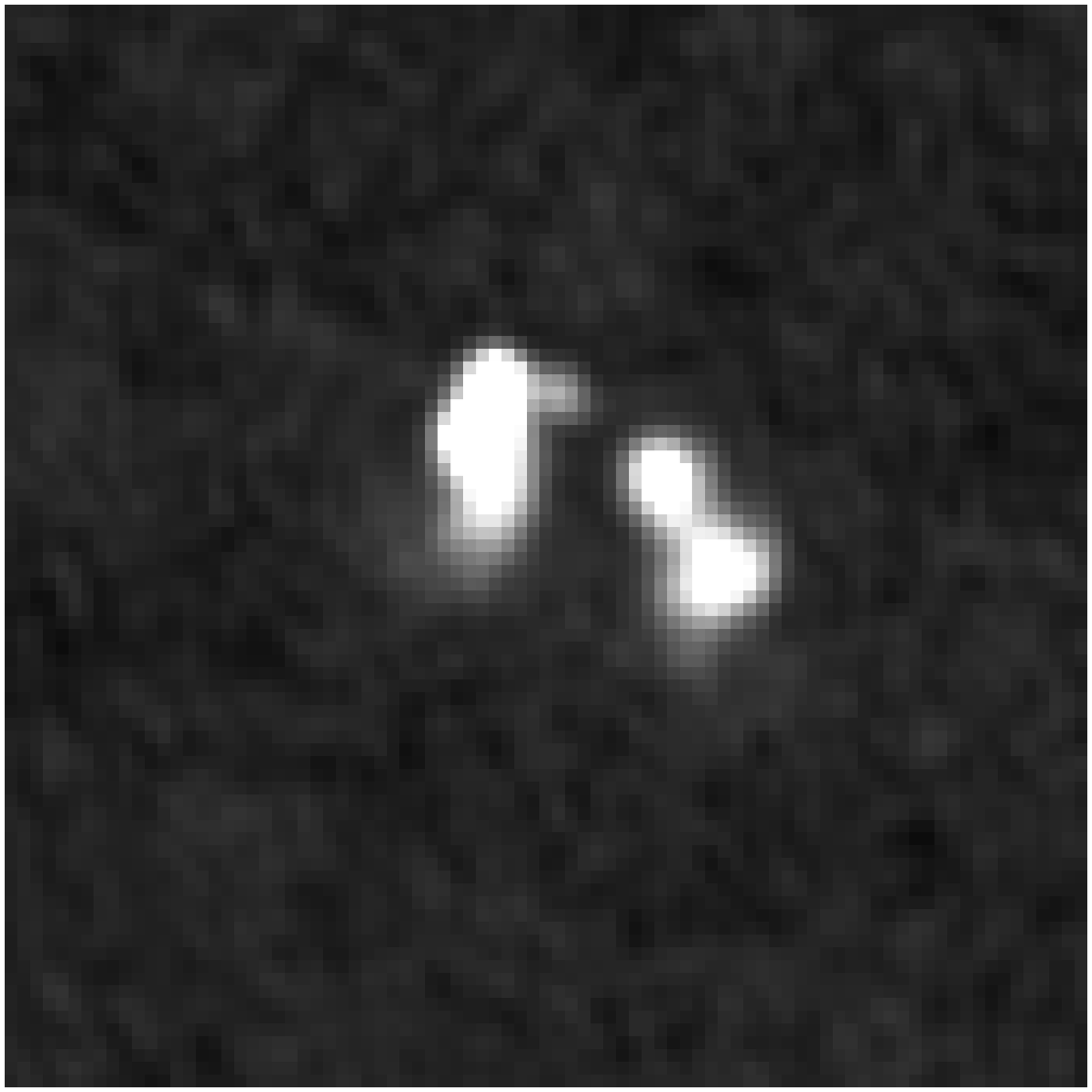}{0.2\textwidth}{(i)}
                    \fig{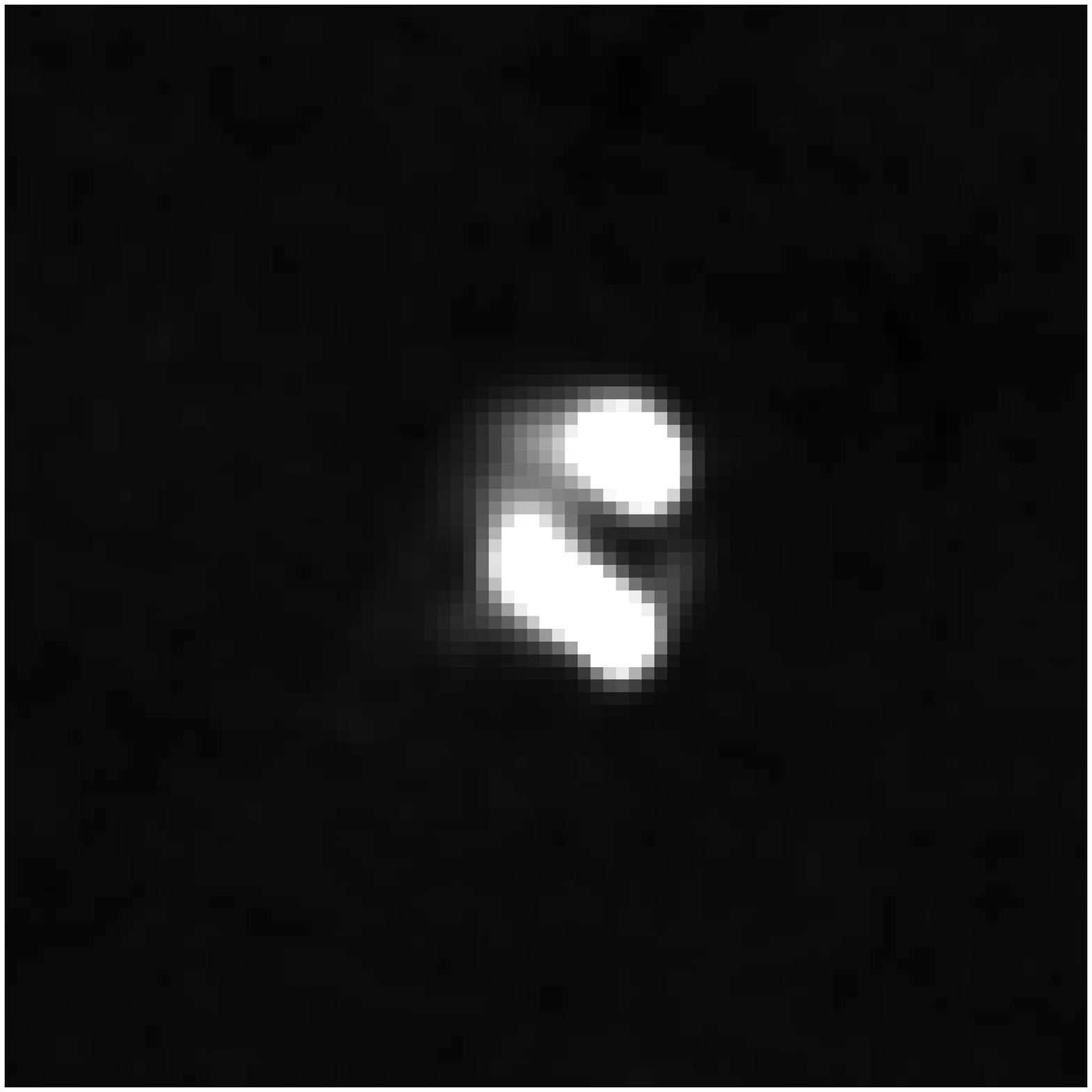}{0.2\textwidth}{(j)}
                    \fig{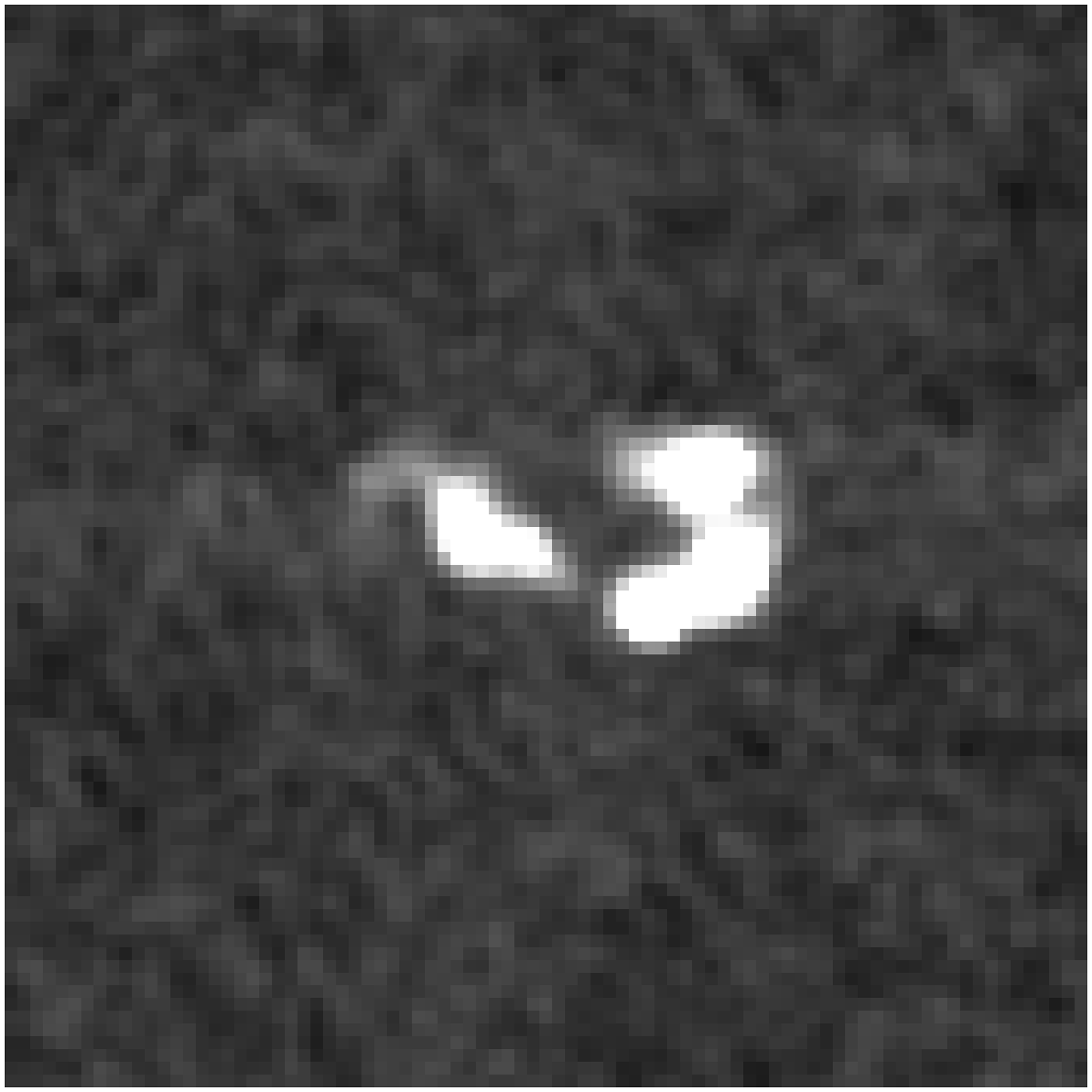}{0.2\textwidth}{(k)}
 \fig{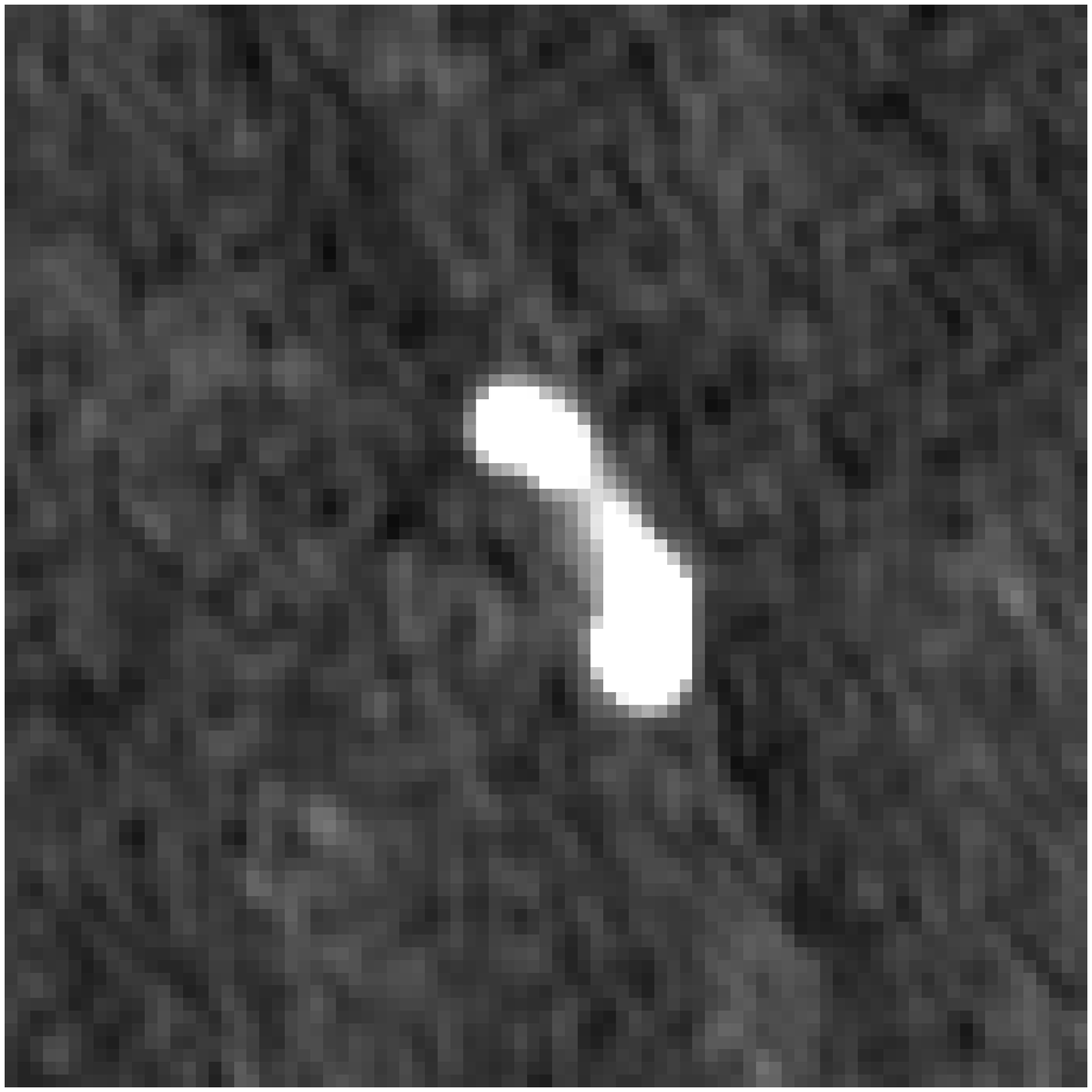}{0.2\textwidth}{(l)}}
\gridline{\fig{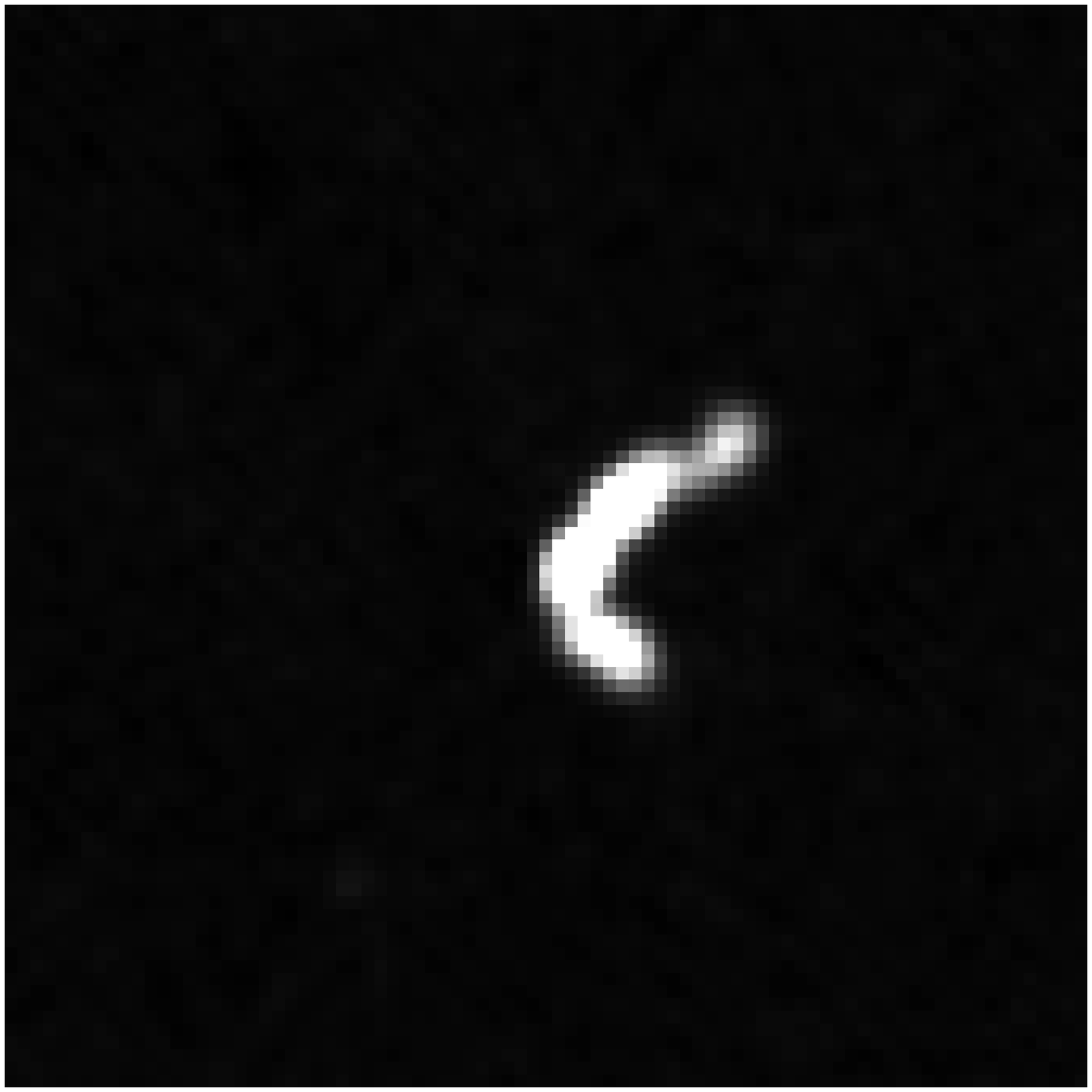}{0.2\textwidth}{(m)}
                   \fig{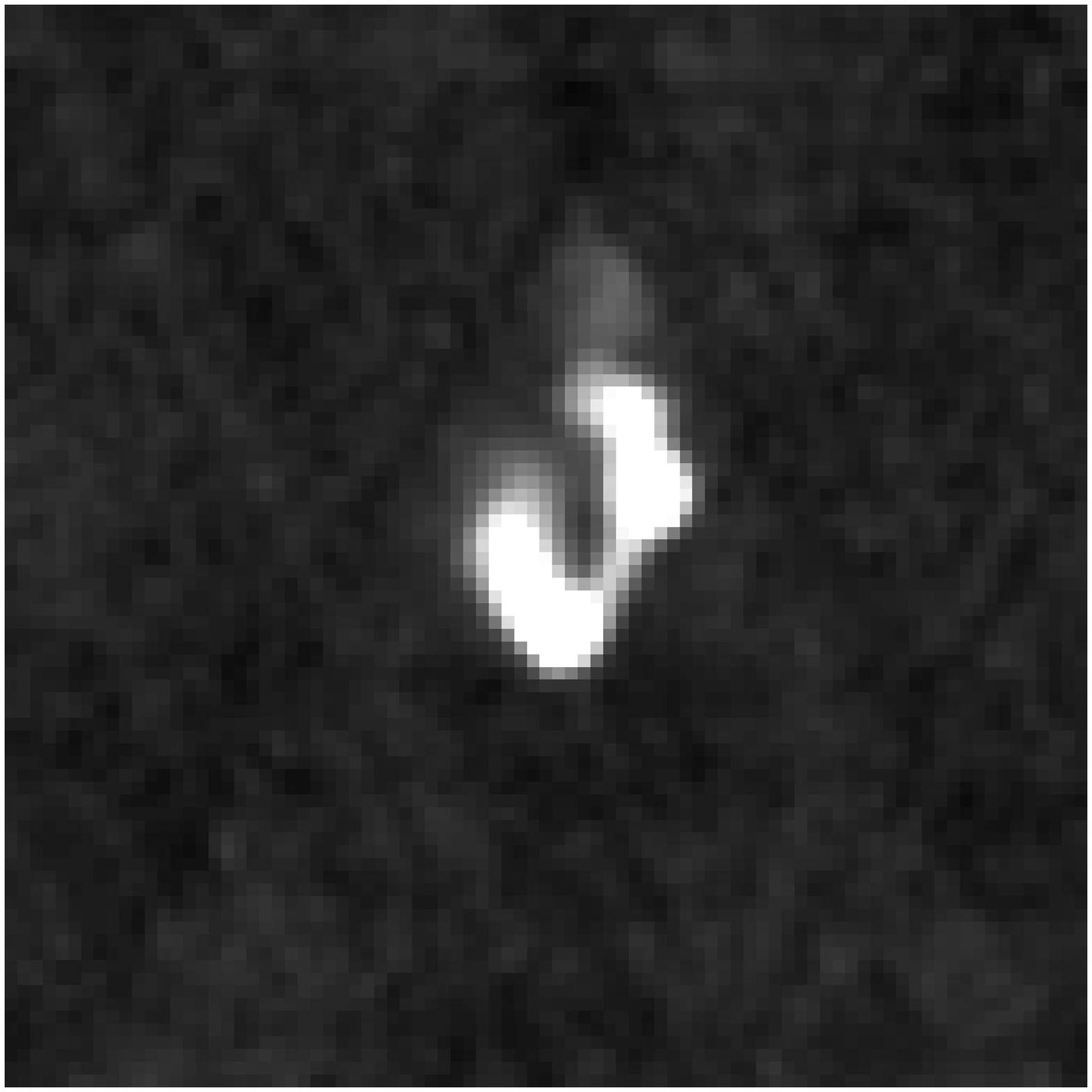}{0.2\textwidth}{(n)}
\fig{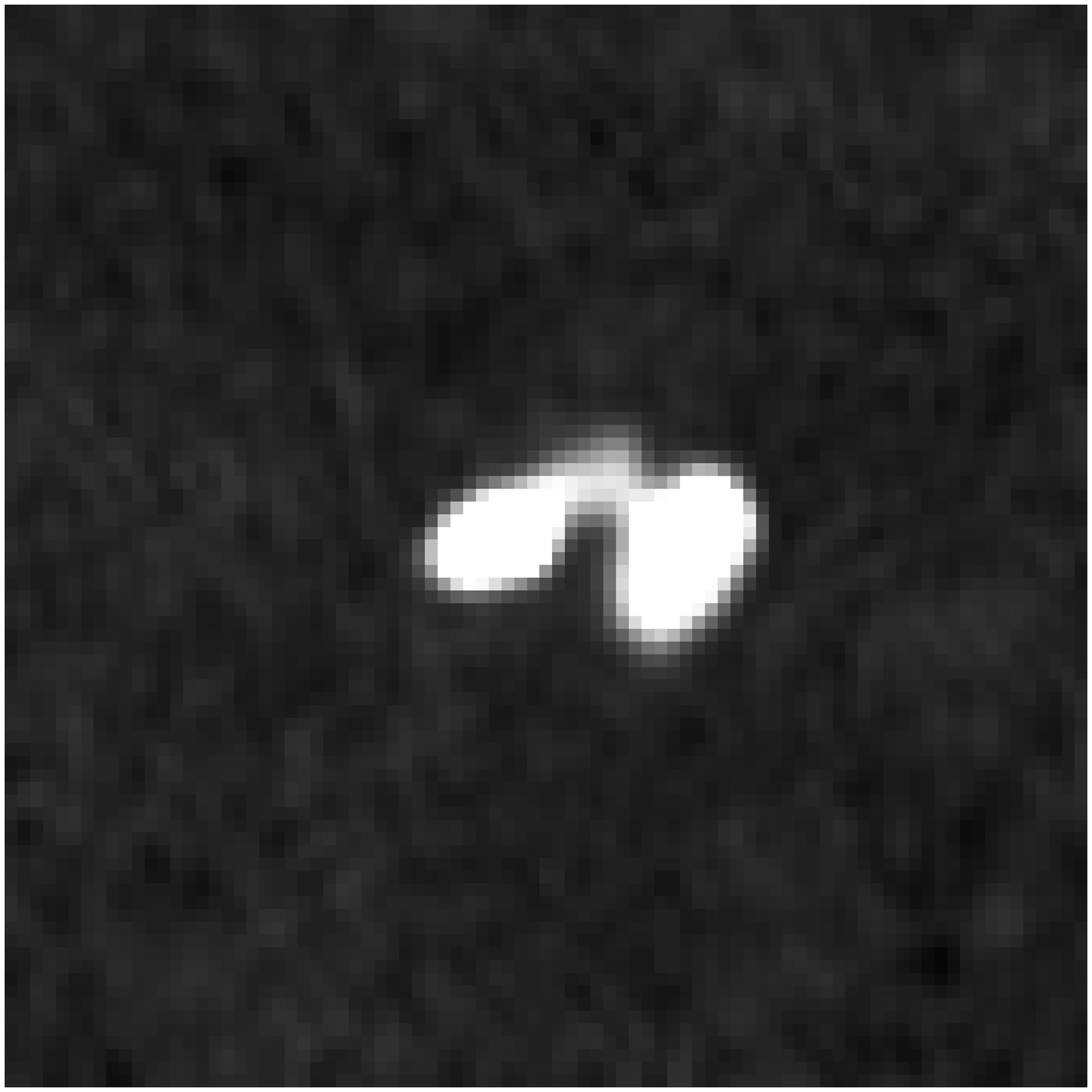}{0.2\textwidth}{(o)}
}
\caption{Bent-tailed radio galaxies misidentified as FR-II type radio galaxies. The prediction result for the validation set showed low precision with high recall for FR-II types radio galaxies and high precision with low recall for Bent-tailed radio galaxies. The figure illustrates the effect of this result with many Ben-tailed radio galaxies misclassified as FR-II radio galaxies.}
\label{bent-misid-frii}
\end{figure}

The FRII radio galaxy classifications have excellent recall at 91\%, but poorer precision at 75\% compared to the other two classes. Since the FRI classifications have both high recall and precision, the precision for FRII classification can be directly linked with the recall of bent-tailed sources.  This implies that sources which are being identified as FRII are actually bent-tailed radio galaxies. Figure \ref{bent-misid-frii} shows these sources for our validation sample, many of these showing two or more bright spots. It may be possible that the algorithm is confused by the bright spots and the diffuse emission in these sources did not get the same 'weight', leading to the misclassification as FRII radio galaxies. 

Overall, the results are comparable to manual classification, while being many times faster. This technique, when applied in an iterative manner would likely reduce the misidentification rate (as seen in FRII-bent classification), with increasing sample size available for training. The effect of training sample size is shown in Table ~\ref{relativeresults}. To do this, we chose a training sample which was $25\%$ of the total training sample and created a classification model with it. The validation sample remained the same across the three classification models. The next training sample was obtained by \textit{incrementing} this sample by a factor of two and generating a classification model with the new training sample. The results show that the average precision, recall and F1 scores, all improve with increasing training sample size. 

\begin{table}[!htb]
\centering
\begin{tabular}{ccccccc}
\hline
\multirow{2}{*}{\textbf{Class}} & \multicolumn{2}{c}{\textbf{Training Samples}} & \multirow{2}{*}{\textbf{Precision (\%)}} & \multirow{2}{*}{\textbf{Recall(\%)}} & \multirow{2}{*}{\textbf{F1-Score(\%)}} & \multirow{2}{*}{\textbf{Support}} \\ \cline{2-3}
                                & \textbf{Actual}      & \textbf{Augmented}      &                                          &                                      &                                        &                                   \\ \hline \hline
Bent-tailed                     & 177                  & 25488                   & 95                                       & 79                                   & 87                                     & 77                                \\ \hline
FR I                            & 125                  & 36000                   & 91                                       & 91                                   & 91                                     & 53                                \\ \hline
FR II                           & 227                  & 32688                   & 75                                       & 91                                   & 83                                     & 57                                \\ \hline
Average                         & \multicolumn{2}{l}{}                          & 88                                       & 86                                   & 86                                     & 187                               \\ \hline
\end{tabular}
\caption{The table shows the class of the source, size of the training samples for each class, Precision, Recall and F1-score of classification for the validation sample as well as the support.}
\label{results}
\end{table}

\begin{table}[!htbp]
\centering
\begin{tabular}{cccc}
\hline
\multicolumn{1}{c}{\textbf{Relative Sample Size}} & \multicolumn{1}{c}{\textbf{Avg Precision(\%) }} & \multicolumn{1}{c}{\textbf{Avg Recall (\%)}} & \multicolumn{1}{c}{\textbf{Avg F1 score(\%)}} \\ \hline \hline
25 \%                                               & 54                                          & 50                                       & 51                                         \\ \hline 
50 \%                                               & 66                                          & 65                                       & 65   \\ \hline                                      

\end{tabular}
\caption{The table shows the results of variation in the training sample size. The first column shows the training sample size relative to the complete sample described in Table ~\ref{results} and the other three columns give the respective weighted averages of precision, recall and F1 score. }
\label{relativeresults}

\end{table}

Table \ref{valtable} shows some of the predictions with probability for the validation samples with their true class and their coordinates.

\begin{deluxetable}{cccccc}
\tablecaption{Table of predictions for validation samples \label{valtable}}
\tablehead{
\colhead{Source} & \colhead{RA} & \colhead{DEC} & \colhead{True Class} & \colhead{Prediction} & \colhead{Probability} \\
\colhead{} & \colhead{h:m:s} & \colhead{d:m:s} & \colhead{} & \colhead{} & \colhead{}\\
}

\startdata
1426+0093           & 14 26 49.84  &  +00 55 59.9   & FRII  & FRII & 99.99995 \\   
3C 194              & 08 10 03.67  &  +42 28 04.0   & FRII  & FRII & 99.99995 \\   
3C 208              & 08 53 08.83  &  +13 52 55.3   & FRII  & FRII & 99.99735 \\   
3C 228              & 09 50 10.77  &  +14 19 57.3   & FRII  & FRII & 99.9999  \\   
3C 240              & 10 17 49.77  &  +27 32 07.7   & FRII  & FRII & 99.99785 \\   
3C 243              & 10 26 31.96  &  +06 27 32.7   & FRII  & FRII & 100.0    \\   
3C 244.1            & 10 33 33.87  &  +58 14 37.9   & FRII  & FRII & 99.9998  \\   
3C 251              & 11 08 37.60  &  +38 58 42.1   & FRII  & FRII & 99.9981  \\   
3C 268.2            & 12 00 59.77  &  +31 33 57.9   & FRII  & FRII & 99.996   \\   
3C 268.4            & 12 09 13.52  &  +43 39 18.7   & FRII  & FRII & 99.99425 \\   
3C 277.2            & 12 53 32.70  &  +15 42 27.3   & FRII  & FRII & 99.98255 \\   
3C 294              & 14 06 44.10  &  +34 11 26.2   & FRII  & FRII & 99.99975 \\   
3C 322              & 15 35 01.27  &  +55 36 49.8   & FRII  & FRII & 99.99985 \\   
3C 323.1            & 15 47 44.23  &  +20 52 41.0   & FRII  & FRII & 99.9888  \\   
3C 336              & 16 24 39.42  &  +23 45 17.5   & FRII  & FRII & 99.99995 \\   
3C 342              & 16 36 37.38  &  +26 48 06.6   & FRII  & FRII & 99.9957  \\   
4C -00.55           & 14 23 26.70  &  -00 49 56.5   & FRII  & FRII & 99.99925 \\   
4C 01.39            & 13 57 01.51  &  +01 04 39.7   & FRII  & FRII & 99.9999  \\   
4C 03.21            & 11 11 22.71  &  +03 09 10.4   & FRII  & FRII & 86.14455 \\   
4C 05.53            & 11 48 47.51  &  +04 55 27.7   & FRII  & FRII & 99.99995 \\   
J151056.2+054441    & 15 10 55.851 &  +05 44 39.29  & BT    & FRII & 99.95345 \\   
J151744.96+310015.8 & 15 17 44.96  &  +31 00 15.8   & FRI   & FRI  & 99.9999  \\   
J152439.9+620225    & 15 24 42.006 &  +62 02 50.93  & BT    & BT   & 99.9971  \\   
J152522.33+314037.1 & 15 25 22.33  &  +31 40 37.1   & FRI   & FRI  & 99.99995 \\   
J153522.1+342247    & 15 35 22.994 &  +34 23 02.98  & BT    & BT   & 99.9821  \\   
J153616.2+142045    & 15 36 16.805 &  +14 20 41.16  & BT    & BT   & 99.6882  \\   
J153932.09+013710.5 & 15 39 32.09  &  +01 37 10.5   & FRI   & FRI  & 100.0    \\   
J154549.4-024954    & 15 45 48.671 &  -02 49 59.76  & BT    & BT   & 99.99845 \\   
J155222.36+223311.9 & 15 52 22.36  &  +22 33 11.9   & FRI   & FRI  & 99.9942  \\   
J155721.38+544015.9 & 15 57 21.38  &  +54 40 15.9   & FRI   & FRI  & 99.9996  \\   
J160318.6+192414    & 16 03 18.856 &  +19 24 18.13  & BT    & BT   & 99.97035 \\ 
\enddata

\end{deluxetable}

\begin{figure*}[ht!]
\centering
\gridline{\fig{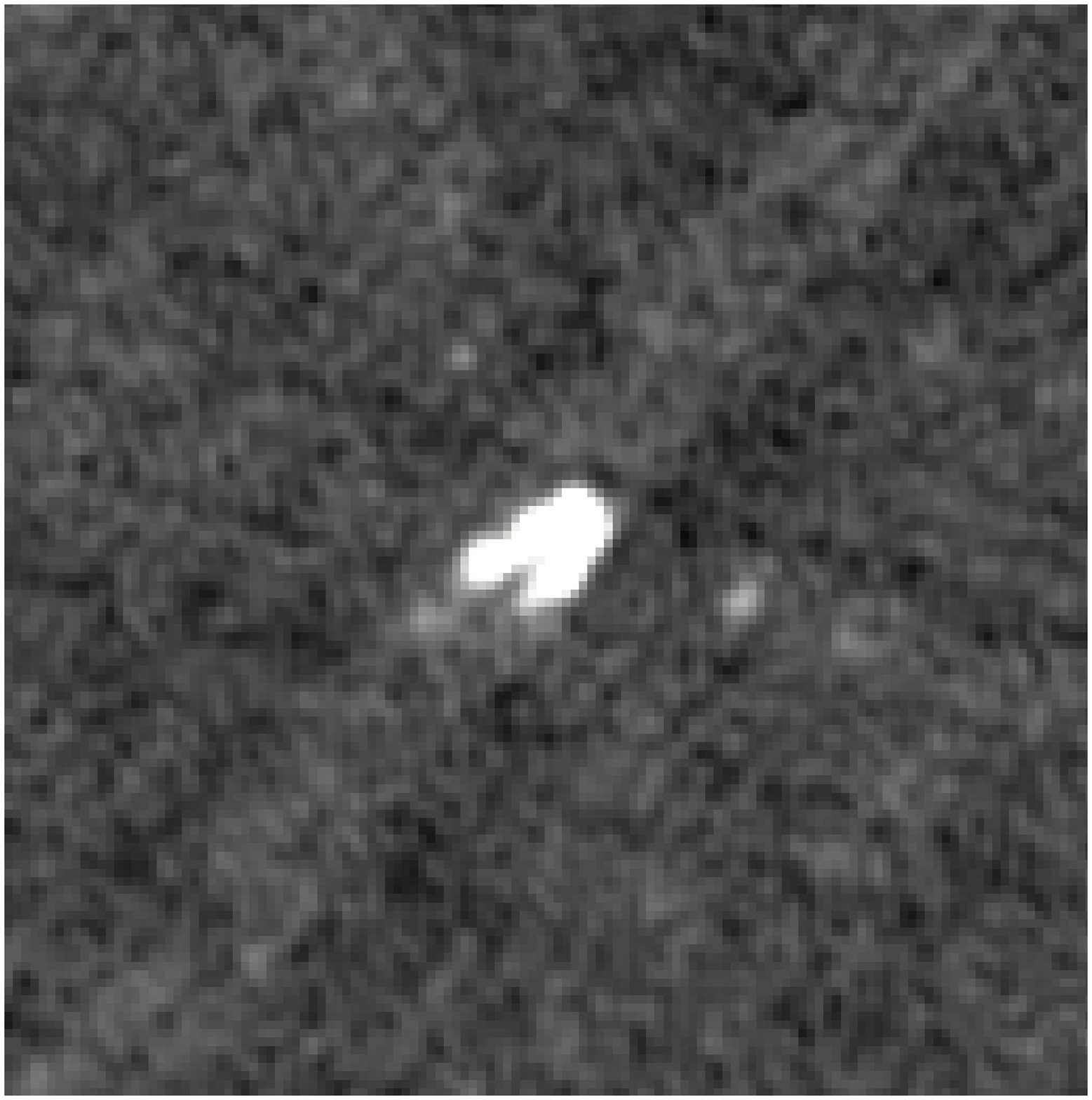}{0.25\textwidth}{(a)}
\fig{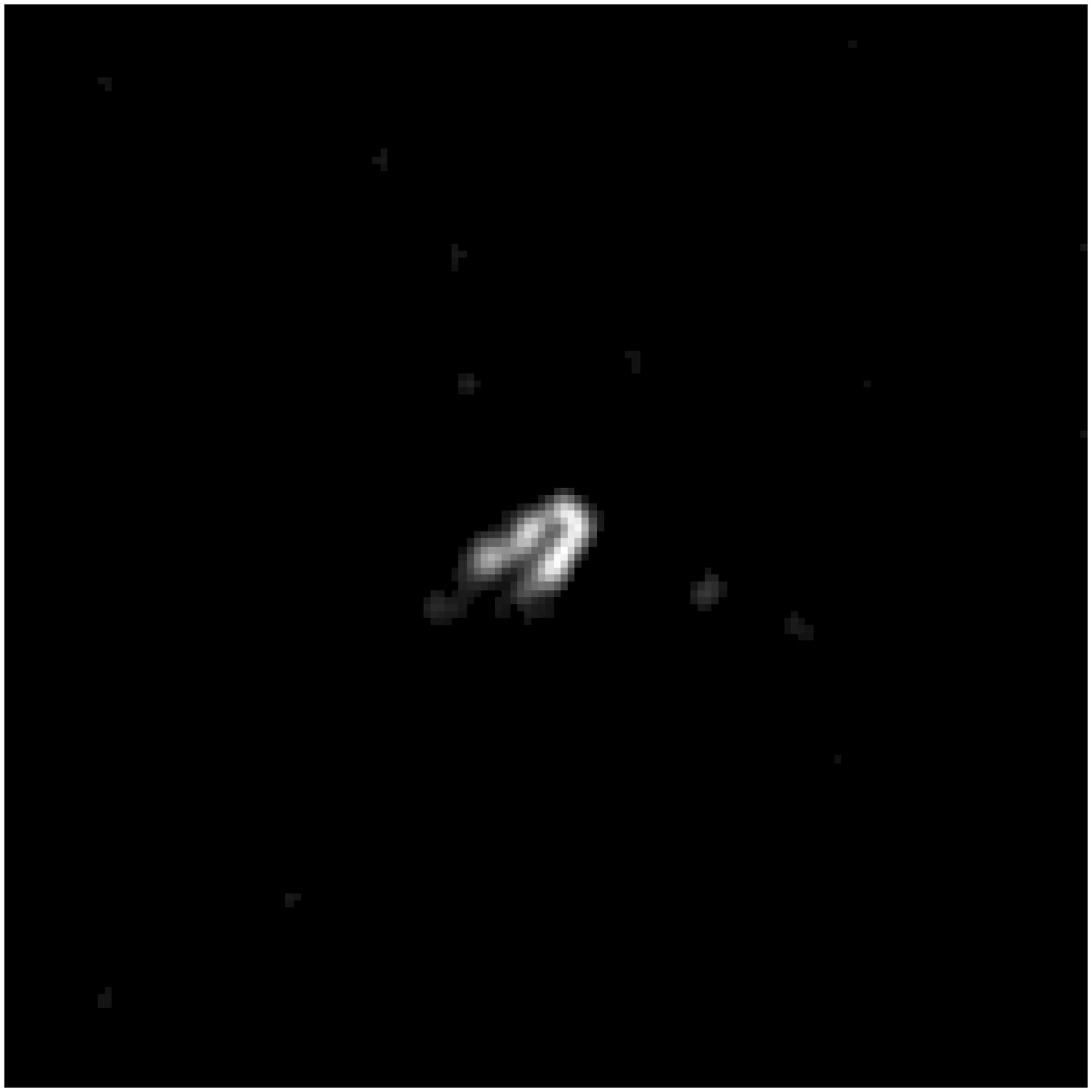}{0.25\textwidth}{(b)}
}
\caption{Sample (J163401.9+062637) from the validation set (a) without any preprocessing and (b) after sigma clipping. The sigma clipped image on the left has far less artefacts and background noise compared to the raw image on the left. This shows the effect of preprocessing the images before being fed into the classifier. The sample without the sigma clipping was incorrectly classified during the validation process.}
\label{preeffects}
 \end{figure*}

One observation that we found during the study was that the convolutional neural network was very sensitive to the preprocessing done to the images. During the training of the network, we performed sigma-clipping of the images before feeding them to the network. The same procedure has to be done for predictions with the network. Figure \ref{preeffects} shows validation sample J163401.9+062637 before and after preprocessing.  

In the example shown in Figure \ref{preeffects}, the sample image was incorrectly classified without sigma clipping and was correctly classified with high confidence after the preprocessing. In this case the actual class label was bent-tail radio galaxy and the prediction without preprocessing was FRII. Depending on the resolution and noise statistics of any image, sigma-clipping can have slight effects on the final image which can also affect the predictions. 

\section{Wider Application of Deep learning model}
Machine learning algorithms trained on data from a specific survey has to be retrained to be used on data from other surveys. Shallow machine learning algorithms need to be retrained from scratch for this purpose which is not realistic in the case of radio astronomy, mainly due to the limitation of sufficient labeled training data. Deep learning methods also suffer from the issue of the need to be retrained, however deep neural networks especially DCNNs like the one in this work need not be trained from scratch. The idea of transfer learning discussed in Section \ref{convolutional_neural_network} makes it possible to use an already trained network model to be retrained with fewer examples from a different survey. 

The main idea of transfer learning in the context of radio galaxy morphological classification can be explained as follows. The initial layers of the neural network will have learned the basic features like edges and bright spots of the input data. The complicated features are always learned in the last few layers. So in the case of classifying radio galaxies, the initial layers of the network learns the basic shape related features of the different radio galaxies. From Figure \ref{filter_visualization} it is evident that the initial layers of the network has learned the basic features of the radio galaxies. For the three different morphologies discussed in this paper, the basic features will be relatively same irrespective of the survey. The last few layers which learns more complex features would be dependent on the resolution and other factors which differ with surveys. Therefore it is possible to retrain the network to work on images from other surveys by retraining only the last few layers and freezing the initial layers. 

There are different variations and methodologies of transfer learning. The methodologies are designed and optimized for various applications. For different methodologies and applications, the number of training samples required for retraining a network will be different. With some methodologies discussed in Section \ref{convolutional_neural_network} the required number of training samples needed to retrain with new dataset is less compared to the original number of samples that was used to train the model. The number of the training samples required are in the order of 1000s for normal image processing applications, however this has not been tested for any astronomical applications. All applications of transfer learning found in literature are done with standard imaging datasets specifically designed for computer vision applications with high signal-to-noise ratio. Since the signal-to-noise ratio of astronomical images are not comparable to those imaging applications, the numbers associated with the training samples may slightly differ with astronomical images.

A demonstration of the use of transfer learning is beyond the scope of this study. This is because  even though the idea of transfer learning is simple, the implementation needs a thorough and systematic analysis as there is no preexisting study that discusses optimizing transfer learning for radio image data. Transfer learning and fine tuning the network for other surveys depend on many factors such as the number of samples in the new dataset, selection of layers that need to be retrained, size of the layers and learning rate. The network may be prone to over-fitting depending on the size of the new dataset and content. There is no clear guideline on which of the initial layers to be frozen and optimized specifically for radio images. The assumption with an already trained network is that it has learned the classification problem with high accuracy (above 90\%). Therefore retraining will be done with smaller learning rates. But if the accuracy is below certain thresholds, this rule will not hold true. A detailed study on optimizing the implementation of transfer learning for radio images is ongoing and will be published in another paper.

Even though these challenges exist, the model that we present here enables astronomers to use for not only classification purposes but also other applications with data from other surveys. With upcoming telescopes, this will enable easy integration of the automated classification system to their science processing pipelines.

\section{Conclusions}
\label{conclusions}
To summarize, in this study, we demonstrate the utility of Machine Learning Techniques in handling large datasets by using deep neural networks to classify images of extended radio galaxies. We use archival data from the FIRST radio survey to train as well as test a convolutional neural network. Initial samples of $\sim 150-200$ sources were used for each class, augmented by rotated versions of these images to train the network. We test the resulting model on a separate validation sample. The results show that the derived model displays good performance across the source categories which we have examined. We find that the precision is highest for the bent-tailed radio galaxies, at $95\%$, whereas it is $91\%$ and $75\%$ respectively for FRI and FRII classes. The recall is highest for the FRI/II classes at $91\%$ and is at $79\%$ for bent-tailed radio galaxies. These results show that the neural networks can reliably identify different classes of radio galaxies, and are comparable to manual classification, while being much faster, and are thus a good technique for source classification and identification when dealing with large image-based datasets. 

At present deep learning techniques are performing with unprecedented accuracies for different classification problems. Bringing these techniques to radio astronomy is critical for handling the data from upcoming radio facilities such as the SKA and its precursors. Early methods involving pattern recognition and shallow machine learning methods are mainly dependent on hand crafted features, which may not completely capture the properties of the radio galaxies. Our methods with DCNN completely removes the layer of handcrafted feature extraction and builds an end-to-end machine-based model. This method completely embraces the principle of \textit{learning from data} and is a novel approach in radio astronomy.  Another consideration is that of the processing time. The time required to classify a single image with this model is less than 0.17 seconds. Even though the classification is very quick, the inference time for convolutional neural network can be further improved with faster GPUs, and by changing the batch size of the input.

Some of the issues we have identified which pertain to radio astronomy data as well as the specific methods employed, are as follows. One of the main requirements and disadvantages of deep learning models is the large sample size required for training. The level of precision obtained with the present model is mainly dependent on the size of the training samples. Hence, large training samples are essential for the use of  'supervised' machine learning methods. Here we have tried to solve the issue by 'bootstrapping' the available images to generate a semi-synthetic dataset. However, this may result in a smaller feature space for the neural network to run and results for datasets not originating from the same observations may suffer. Yet another issue is that the the techniques used show heavy dependence on pre-processing. With images from different surveys the pre-processing will affect the inference of the classifier. Developing machine learning techniques to make inferences in non-stationary environments is still an open problem.

Another issue that affects the quality of the trained model is the number of representative samples that each class originally had. We originally had fewer FRI radio galaxy samples compared to FRII and Bent-tailed radio galaxies. Even though the `bootstrapping' generated enough samples to train the network, the representative samples for each class were different. Therefore the features learned during the training will be confined for each class in the feature space making the model less general and in turn reducing the overall accuracy. We tried to push the accuracy limits of the model by generating the synthetic samples and modifying the loss functions and the success was limited. Since the model allows for transfer learning, this issue can be managed by retraining the model with new samples from future catalogs.

We aim to make the code and model publicly available to the community. The Caffe model and associated code for classification will be available in public domain at https://github.com/ArunAniyan/RadioGalaxyClassification. An online web service which permits a radio image to be uploaded for classification is also under construction. This will also help improving the model with feedback from the users and by retraining with more samples, enabling the astronomers to use the service for research purposes with better accuracy.

\section{Acknowledgments}
\label{acknowledgements}
We thank the Square Kilometer Array South African Project (SKA SA), the SKA SA postgraduate bursary program and the South African Research Chair Initiative (SARChI) program for funding the research project. This research has been conducted using resources provided by the Science and Technology Facilities Council (STFC) through the Newton Fund and the SKA Africa. We thank the anonymous referee for the comments and suggestions which have improved the manuscript considerably. We would like to also thank Prof.Oleg Smirnov, Etienne Bonnassieux, Dr.Nadeem Oozeer and Dr.Jasper Horrell for their valuable suggestions and comments. We also thank Dr.Lindsay Magnus for his inputs on the MeerKAT data rates. The authors would also like to thank Dr.Roger Deane for detailed feedback which was instrumental to this work.  
\bibliography{reference}

\end{document}